\documentclass[reprint,superscriptaddress,amsmath,amssymb,aps,preprintnumbers]{revtex4-1}
\usepackage{graphicx}
\usepackage{caption}
\usepackage{dcolumn}
\usepackage{bm}
\usepackage[usenames]{color}
\usepackage[normalem]{ulem}
\RequirePackage{graphicx,epstopdf}
\usepackage[colorlinks=true, pdfstartview=FitV, linkcolor=blue, citecolor=blue, urlcolor=blue]{hyperref}

\renewcommand{\sout}[1]{\bgroup \color{red} \ULdepth=-.5ex \ULset {#1}}
\newcommand{\comment}[1]{}
\graphicspath{{./Figs/}}
\newcommand{\refeq}[1]{(\ref{#1})}
\newcommand{\average}[1]{\ensuremath{\langle#1\rangle}}
\newcommand{\xc}{ x_{\rm c} }
\newcommand{\ttau}{ \tilde{\tau} }
\newcommand{\teta}{ \tilde{\eta} }

\newcommand{\bx}{ \bm{x} }
\newcommand{\bk}{ \bm{k} }
\newcommand{\bb}{ \bm{b} }
\newcommand{\bxp}{ \bm{x}_\perp }
\newcommand{\bkp}{ \bm{k}_\perp }
\newcommand{\bbp}{ \bm{b}_\perp }

\newcommand{\xm}{ x^- }
\newcommand{\xp}{ x^+ }
\newcommand{\xmp}{ x^{\mp} }

\newcommand{\snn}{ \sqrt{s_{\rm NN}} }
\newcommand{\qs }{ Q_{\rm s} }
\newcommand{\qsn}{ Q_{\rm s, n} }
\newcommand{\qsa}{ Q_{\rm s, A} }
\newcommand{\imp}{ b_{\rm imp} }
\newcommand{\no}{ N_{\rm 1D} }

\newcommand{\rl}{ r_{\rm L} }
\newcommand{\cl}{ l_{\rm L} }
\newcommand{\sigl}{ \sigma_{\rm L} }
\newcommand{\uv}{ \Lambda_{\rm UV} }
\newcommand{\ir}{ \Lambda_{\rm IR} }
\newcommand{\elrf}{ \varepsilon_{\rm LRF} }

\newcommand{\ec}{ \varepsilon_{\rm c} }

\begin{document}
\preprint{}
\title{
Simulation of a (3+1)D glasma in Milne coordinates: 
Topological charge, eccentricity, and angular momentum
}
\author{Hidefumi Matsuda}
\email{da.matsu.00.bbb.kobe@gmail.com}
\affiliation{Zhejiang Institute of Modern Physics, Department of Physics, Zhejiang University, Hangzhou, 310027, China}
\author{Xu-Guang Huang}
\email{huangxuguang@fudan.edu.cn}
\affiliation{Physics Department and Center for Particle Physics and Field Theory, Fudan University, Shanghai 200438, China}
\affiliation{Key Laboratory of Nuclear Physics and Ion-beam Application (MOE), Fudan University, Shanghai 200433, China}
\affiliation{Shanghai Research Center for Theoretical Nuclear Physics, National Natural Science Foundation of China and Fudan University, Shanghai 200438, China}

\begin{abstract}
We apply the 3D glasma simulation method using Milne coordinates, proposed in our previous work~\cite{3Dglasma_FU1}, 
to the early stage of the Au-Au collisions at $\snn=200$ GeV. 
The nucleus model prior to the collisions, which offers the initial condition for the 3D glasma simulation 
is constructed to account for the longitudinal structure of the nucleus, 
the finite thickness of nucleons and their random positions along the collision axis.
We investigate rapidity profiles for a wide range of physical quantities of the glasma, 
including energy, pressure, fluctuations of topological charge, eccentricity, and angular momentum. 
In particular, we elucidate the behavior of eccentricity and angular momentum, 
which are physical quantities dependent on the geometric shape of the glasma, across a wide range of impact parameter regions.
\end{abstract}
\maketitle
\section{Introduction}\label{Sec:0}
Relativistic heavy-ion collisions provide a unique opportunity to study the properties of matter under extreme conditions.
Understanding the evolution of the matter produced in these experiments is a complex problem involving various aspects of quantum chromodynamics (QCD).
In the early stages of the collision, the system is energetic and consists of quarks and gluons which are liberated from an inside of a hadron.
This matter is in a far-from-equilibrium state, and is expected to evolve into a quark-gluon plasma (QGP), where quarks and gluons behave collectively as a hydrodynamic fluid.
As the system expands and cools, the QGP undergoes a phase transition to a many-body system of hadrons; however, it can still be regarded as a hydrodynamic fluid.
Further expansion leads to the breakdown of the hydrodynamic description due to the dilution of the system, and then kinetic theory becomes valid to describe the dilute hadron gas.

To confirm the temporal existence of the QGP and study its features, it is necessary to accurately understand the entire evolution of the produced matter, noting that the dominant dynamics and suitable theoretical frameworks differ at each stage.
The collective behavior from the formation of the QGP to the early stage of the hadron gas is well-described by relativistic viscous hydrodynamics~\cite{Hydro_for_HIC}, and the evolution of the dilute hadron gas is effectively captured by hadron transport models, such as the Ultra-relativistic Quantum Molecular Dynamics (UrQMD)~\cite{UrQMD1998,UrQMD1999} and the Jet $AA$ Microscopic Transportation Model (JAM)~\cite{JAM}.
However, understanding non-equilibrium matter prior to the hydrodynamical stage remains inadequate because it involves complex non-equilibrium processes in QCD, from the formation of far-from-equilibrium matter due to the collision to the hydrodynamization of the produced matter.
In practice, instead of pursuing the full description for the non-equilibrium process, we adopt an approach that utilizes a phenomenological initial state model to provide the initial conditions for hydrodynamic fluids.

Initial state models can be broadly categorized into two types.
The first type is a non-dynamical model that constructs the profile of the energy or entropy density at the phenomenological onset time of hydrodynamics directly by considering some relevant information, such as the event-by-event geometry of nucleons within the nucleus, and nucleon-nucleon scattering amplitude.
Examples include the Monte Carlo Glauber model~\cite{MonteCarlo}, and Reduced Thickness Event-by-event Nuclear Topology (T\raisebox{-0.5ex}{R}ENTo) model~\cite{TRENTo1,TRENTo2}.
The second type is a dynamical model that focuses on specific degrees of freedom of the non-equilibrium matter and incorporates their pre-equilibrium flow into the hydrodynamic initial conditions. 
This is achieved by simulating the dynamical evolution of these degrees of freedom using an effective description.
For instance, the IP-glasma (impact-parameter-dependent glasma) model provides the hydrodynamic initial condition, where the pre-equilibrium information of the gluonic matter produced in the collision, called glasma, is taken into account~\cite{IPglasma}.

The IP-glasma model utilizes the classical Yang-Mills (CYM) field to describe the glasma.
This CYM field description is based on the color glass condensate (CGC) effective theory, which is an effective theory to describe the collective dynamics of gluons inside a large and relativistic nucleus~\cite{MV,CGCreview}.
In a large nucleus at a relativistic speed, the number of gluons carrying a small fraction of the parent hadron's longitudinal momenta (soft partons) is overwhelmingly large compared to the number of partons carrying a larger fraction of the parent hadron's longitudinal momenta (hard partons).
The high-occupancy of these soft gluons allows them to be approximated as classical fields, similar to a Bose-Einstein condensate, that are radiated from the hard partons.
Since soft gluons do not see individual hard partons, but rather the collection of a lot of hard partons, the color charge seen by soft gluons appears disordered.
In addition to the disordered behavior, the hard partons appear frozen in the lifetimes of the soft gluons due to the Lorentz dilation, together making the hard partons appear glass-like to the soft gluons.
In the CGC effective theory, with the use of these features, a large and relativistic nucleus is modeled as the theoretical state, called color glass condensate, where soft gluons are described as the CYM fields and hard partons are given as the disordered and frozen color charge density that works as a source of the CYM fields.
Based on the CGC framework, relativistic heavy-ion collisions can be modeled as the collision of two CGCs and the associated liberation of soft gluons, using the CYM fields~\cite{MVcollision}.
The CYM description is also valid for the subsequent dynamical evolution of the liberated gluons, the glasma, as far as it is enough dense~\cite{2Dglasma_KV}.
The non-equilibrium dynamics of glasma has been extensively studied based on the CYM field description, under the assumption of boost invariance~\cite{2Dglasma_KV,KNV_SU3,KNV_energy,2Dglasma_L,LM,InstabilityCYM2006_RV,InstabilityCYM2008,InstabilityCYM2009,InstabilityCYM2012FG,InstabilityCYM2012BSSS,InstabilityCYM2013,PressureEG2013,Turbulent2014,Universal2014,Entropy_kyoto,SmallTime_CCM,SmallTime_CCFMP,SmallTime_CM,HP_CPIC}.

While the IP-glasma model is considered one of the most successful initial state models in reproducing experimental results, its conventional application has been limited to the near mid-rapidity region~\cite{IPglasma}.
This limitation is due to the boost invariant assumption for the CYM description.
Recently, experimental measurements have revealed rapidity-dependent observables, such as the rapidity decorrelation of flow~\cite{Dec_flow1,Dec_flow2} and transverse momentum~\cite{Dec_tra1,Dec_tra2}. 
Additionally, the theoretical study suggests that the rapidity correlation of conserved quantities is expected to preserve the pre-equilibrium information encoded in the hydrodynamic initial condition~\cite{Correlation_from_hydro_init_to_obs}. 
These developments inspire us to study the glasma description that goes beyond the boost-invariant assumption to establish a 3D initial state model incorporating the non-equilibrium flow of the glasma.

Two approaches are being explored to account for effects beyond boost invariance in the glasma, each addressing different aspects.
One approach solves the Jalilian-Marian-Iancu-McLerran-Weigert-Leonidov-Kovner (JIMWLK)  equation to account for the next-to-leading order correction to soft gluons inside nuclei prior to the collision, and goes beyond boost invariance~\cite{JIMWLK1,JIMWLK2,JIMWLK3,JIMWLK4,JIMWLK5,JIMWLK6}.
The 3D IP-glasma model that accounts for the next-to-leading order correction has already been established and applied to phenomenology~\cite{3DIPglasma_JIMWLK1,3DIPglasma_JIMWLK2,3DIPglasma_JIMWLK3}.
The other approach is to relax the shock-wave approximation assumed in the boost-invariant glasma description, and the JIMWLK equation.
The so-called shock-wave approximation treats the hard partons in a specific manner: the hard partons are assumed to run at the speed of light, go straight along the collision axis, remain unchanged during the collision, and be located on an infinitely thin nucleus in the collision direction.
Especially, considering the finite thickness of the nucleus breaks the boost invariance.
References~\cite{3Dglasma_FU1,3Dglasma_CPIC,3Dglasma_SS} propose numerical simulation methods for the 3D glasma that relax the latter two of the shock-wave approximations mentioned above.
These simulations solve a set of evolution equations of the CYM field and classical color charge density numerically. 
However, the computational resources required are significantly large, since the longitudinal dimension and the evolution of the classical color charge density must be considered, unlike the calculation of the boost-invariant glasma. 
Therefore, the 3D glasma simulation beyond the shock-wave approximation has not yet reached the stage of being combined with hydrodynamics to compare with experimental results.

The semi-analytical method, which relies on the order-by-order analytic solution of the evolution equations using weak-field expansion and Monte-Carlo integration, helps us to avoid the large numerical demands of the 3D glasma beyond the shock-wave approximation~\cite{3Dglasma_analytic1,3Dglasma_analytic2}. 
This method serves as a valuable guide for understanding the features of the 3D glasma.
However, in realistic heavy-ion collisions, the expected density of the color charge is not weak enough to clearly justify the weak-field expansion. 
Therefore, it is essential to continue investigating the 3D glasma using simulation methods.

Recently, we proposed a new approach to perform the 3D glasma simulation beyond the shock-wave approximation using Milne coordinates, which relatively requires fewer computational resources~\cite{3Dglasma_FU1}.
Previously, the 3D glasma simulation methods beyond the shock-wave approximation have been formulated in Minkowski coordinates, requiring a larger longitudinal lattice size due to the longitudinal expansion of the system~\cite{3Dglasma_CPIC,3Dglasma_SS}.
In contrast, simulations using Milne coordinates that follow the longitudinal expansion of the system significantly alleviate this problem.

In this study, we apply the 3D glasma simulation method in Milne coordinates to Au-Au collisions at 200 GeV per nucleon pair conducted in the Relativistic Large Hadron Collider (RHIC).
We investigate various physical quantities in the 3D glasma that has rapidity dependence due to the consideration of some effects beyond shock-wave approximation.
In Sec.~\ref{Sec:2}, we introduce the CGC-based description of the 3D glasma in Milne coordinates, which is used in our simulations.
In Sec.~\ref{Sec:3}, we construct the 3D initial condition for the gold nuclei prior to the collision, where the finite thickness of nucleons and their random positions along the collision axis are taken into account.
In Sec.~\ref{Sec:4}, we focus on the central collision.
We calculate the rapidity profiles of the energy density and pressure, basic quantities of hydrodynamics, as well as the squared topological charge that is expected to be generated at the moment of the collision.
In Sec.~\ref{Sec:5}, we show some geometry-dependent physical quantities, such as eccentricity, which reflects the geometry in the transverse plane perpendicular to the collision axis, and angular momentum, which is expected to be generated by the asymmetry of the system geometry with respect to the impact parameter direction.
In Sec.~\ref{Sec:6}, we summarize our work.

\section{Method}\label{Sec:2}
We numerically simulate the process of two distant nuclei colliding and separating, followed by the formation of glasma.
In the following, we first introduce two types of Milne coordinates: 
the usual Milne coordinates used to interpret numerical results obtained by simulations 
and the modified Milne coordinates used to describe the evolution of the system in simulations.
Then, we explain the CGC-based description of the 3D glasma simulation in continuous spacetime, based on the modified Milne coordinates. 

\subsection{Usual and Modified Milne Coordinates}

We consider two types of Milne coordinates: one is the Milne coordinates usually used in hydrodynamic simulations, which we call ``usual Milne coordinates," and the other is the Milne coordinates specially introduced, which we call ``modified Milne coordinates."
The former is used to interpret numerical results obtained from simulations, and the latter is used to describe the evolution of the system in simulations.
As preparation for introducing them, we first consider the general expression of the Milne coordinates as
\begin{align}
\tau &= 
\sqrt{2(\xm-x_{\rm shift})(\xp-x_{\rm shift})}\ ,\label{Eq:usu_gen}\\
\eta &=
\frac{1}{2} \ln{\frac{\xp-x_{\rm shift}}{\xm-x_{\rm shift}}} \ ,\label{Eq:usu_gen}
\end{align}
where the light-cone coordinates are defined as $\xmp = (t \mp z)/\sqrt{2}$, and the shift parameter $x_{\rm shift}$ is introduced, which can be regarded as adding a constant to $t$, $t \to t-x_{\rm shift}$.
In our setup, the longitudinal extents of the colliding nuclei are roughly given by their radius $R$ and Lorentz factor $\gamma$, and their center positions in the light-cone variable are taken as $\xm = \xc$ and $\xp = \xc$ with $\xc > R/\gamma$.

\begin{figure*}[tp]
\begin{minipage}{0.45\textwidth}
    \centering
    \includegraphics[width=\textwidth]{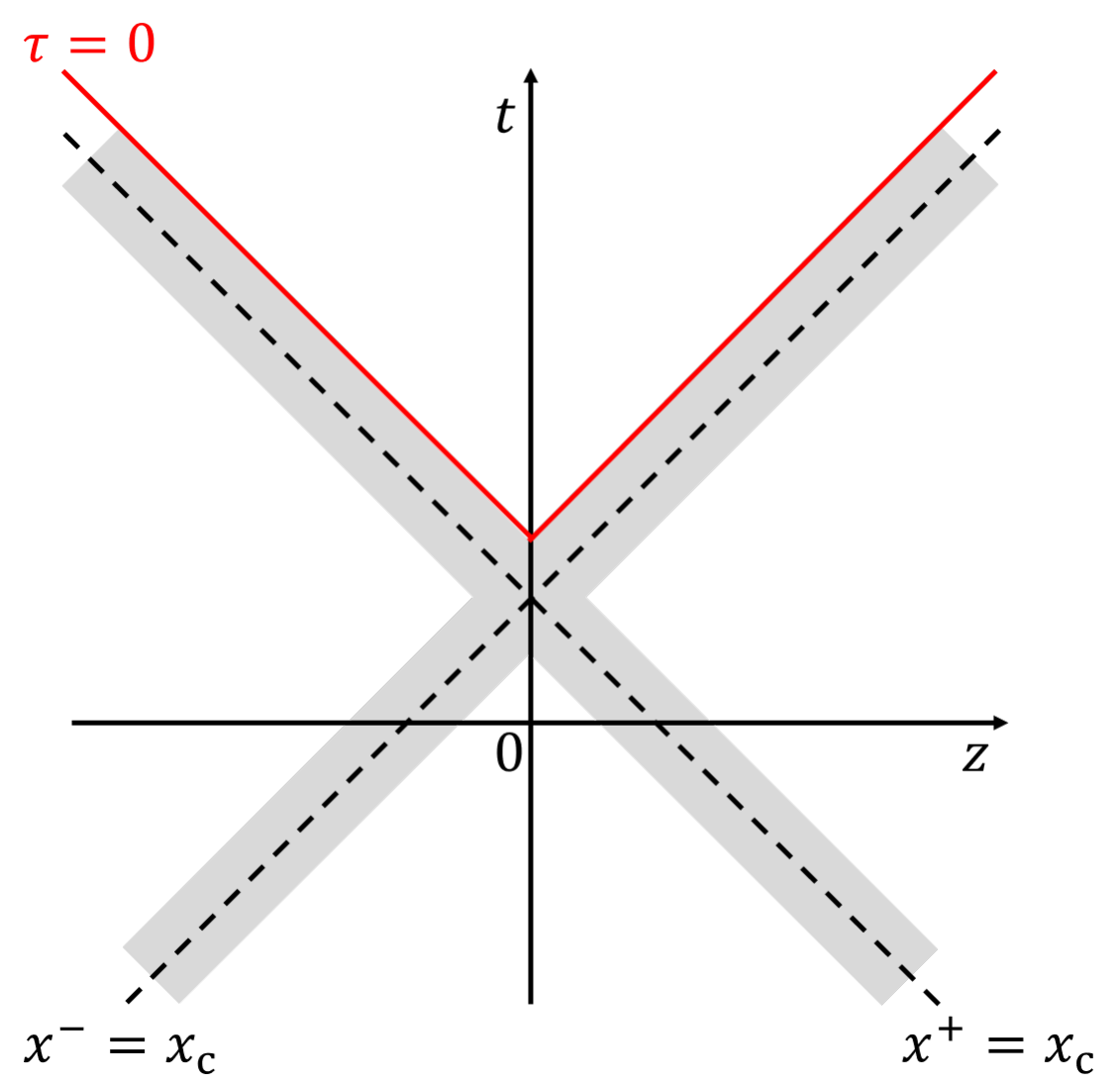}
\end{minipage}%
\hfill
\begin{minipage}{0.45\textwidth}
    \centering
    \includegraphics[width=\textwidth]{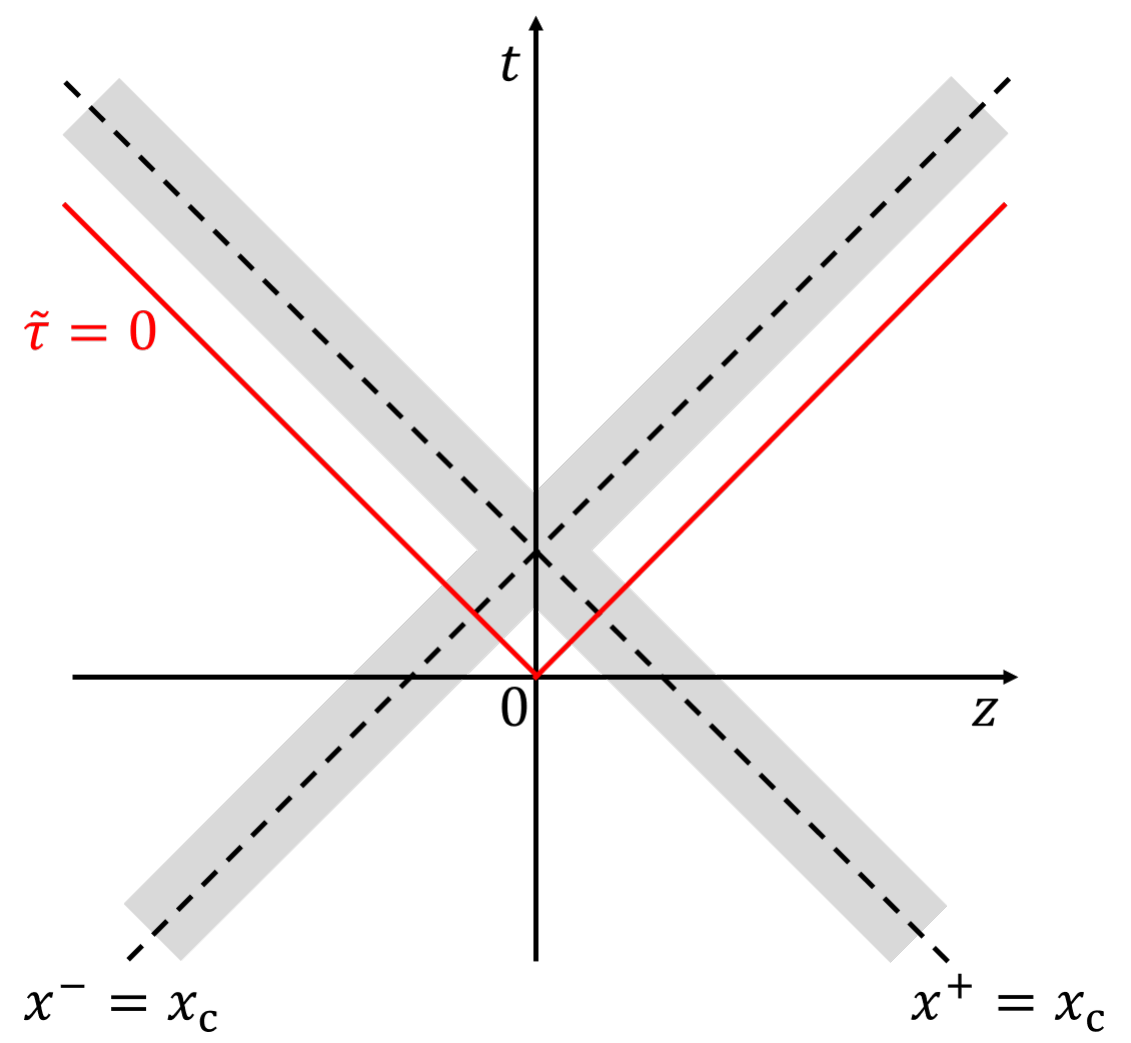}
\end{minipage}
\caption{
Left:  trajectories of the colliding nuclei and $\tau=0$  line on the $t-z$ plane.
Right: trajectories of the colliding nuclei and $\ttau=0$ line on the $t-z$ plane.
The spacetime region of propagating nuclei is colored in gray.
The gray area corresponds to $\xmp = [\xc-R/(\sqrt{2}\gamma),\xc+R/(\sqrt{2}\gamma)]$.
The $\tau=0$ and $\ttau=0$ lines are indicated in red.
}
 \label{Fig:spacetime}
\end{figure*}
In the usual Milne coordinates, the colliding nuclei barely exist in the forward light-cone, as illustrated in the left panel of Fig.~\ref{Fig:spacetime}.
Accordingly, we set $x_{\rm shift} = \xc + \frac{R}{\sqrt{2}\gamma}$ to define the usual Milne coordinates and denote them as ($\tau, \eta$).
The calculation results obtained from the simulation in the modified Milne coordinates are interpreted using the usual Milne coordinates.
To achieve that, we perform a general coordinate transformation on the numerically calculated energy-momentum (EM) tensor and field strength tensor in the modified Milne coordinates to obtain these in the usual Milne coordinates.

In the modified Milne coordinates, the colliding nuclei exist almost entirely in the forward light-cone, as illustrated in the right panel of Fig.~\ref{Fig:spacetime}.
We define the modified Milne coordinates by taking $x_{\rm shift}=0$ to realize this situation, and denote them as ($\ttau, \teta$).
We conduct numerical simulations for the $\ttau$ evolution of the system, which includes the collision and separation of nuclei as well as the generation and evolution of glasma, in modified Milne coordinates.

\subsection{CGC-based description of Glasma}\label{Sec:formulaton_cont}
We employ the classical field approximation based on the CGC effective theory to describe the initial stages of relativistic heavy-ion collisions.
The CGC effective theory is adept at describing the collective behavior of soft gluons, each carrying a small fraction of the longitudinal momentum of the parent nucleon inside a large and relativistic nucleus.
Its key features include modeling the hard partons, which are quarks and gluons with larger longitudinal momentum, using the classical color charge density $\rho$, which acts as a source radiating the soft gluons, while treating the soft gluons as the CYM field.
The CYM field is given as a solution to the equations of motion containing the classical color charge densities,
\begin{align}
[D_\nu,F^{\nu\mu}]=J^\mu_{\rm A}+J^\mu_{\rm B}
=\delta^{\mu +} 
\frac{\rho_{\rm A}}{g}+\delta^{\mu -} \frac{\rho_{\rm B}}{g}
\ ,\label{Eq:conteq}
\end{align}
where $g$ is the strong coupling constant, $\rho_{\rm A}$ and $\rho_{\rm B}$ denote the classical color charge densities of the right- and left-moving nuclei, respectively, and the covariant derivative is defined as $D_\mu \equiv \partial_\mu-igA_\mu$.
The classical color charge densities obey the following continuity equations,
\begin{align}
[D_\mu,J^\mu_{\rm A/B}]=[\partial_\mu-igA_\mu,\delta^{\mu \pm} \frac{\rho_{\rm A/B}}{g}]=0\ ,\label{Eq:current}
\end{align}
which indicates that hard partons, carrying classical color charges, go straight along the collision axis at the speed of light.
Thus, the change in the classical color charge density according to Eq.~\refeq{Eq:current} is only the phase rotation in the color space.
In this framework, the properties of the nucleus that we are interested in are given by the initial condition of classical color charge density at the infinite past, which is randomly assigned according to a probability distribution, $W_{\rm A/B}[\rho_{\rm A/B}]$.
The simplest way to assign the probability distribution is to assume a Gaussian distribution, known as the McLerran-Venugopalan (MV) model~\cite{MV}.
The Gaussian distribution is justified by the central limit theorem if the color charge density can be treated as the superposition of the color charges of a large number of mutually independent hard partons, such as when the color charge density originates solely from the valence quarks of each nucleon inside a significantly large nucleus~\cite{KV}.
Therefore, as the Bjorken $x$ of the soft gluons of interest and atomic number of the nucleus are larger, the applicability of the MV model is enhanced.
In the MV model, the expectation values of physical quantities in the glasma are given by the event average weighted by the probability distribution for two colliding nuclei,
\begin{align}
\average{\mathcal{O}} = \int \mathcal{D}\rho_{\rm A} \mathcal{D}\rho_{\rm B} 
W_{\rm MV}[\rho_{\rm A}] W_{\rm MV}[\rho_{\rm B}] 
\mathcal{O}[\rho_{\rm A},\rho_{\rm B}]\ .\label{Eq:EV}
\end{align}
Note that, when moving to a larger energy scale or a wider rapidity region, the relevant value of Bjorken $x$ of the soft gluons becomes smaller, and the higher-order correction to the weight function obtained by solving the JIMWLK equation needs to be considered~\cite{JIMWLK1,JIMWLK2,JIMWLK3,JIMWLK4,JIMWLK5,JIMWLK6}.
Then, it is a more complicated problem whether the factorization shown in Eq.~\refeq{Eq:EV} holds under the evolution of the JIMWLK equation, which is discussed in Ref.~\cite{JIWMLK_fact}.

Let us consider the case of a single nucleus without a collision as an introduction.
The solution to the equation of motion for the CYM fields for the moving nucleus is a non-Abelian gauge theory version of the Weizs$\ddot{\rm a}$cker-Williams (WW) field in the electromagnetic field~\cite{MV,WW_QED},
\begin{align}
A^{\rm A/B}_\pm = 0\ ,\ \ 
A^{\rm A/B}_{i=1,2} = \frac{i}{g} 
V_{\rm A/B} \partial_i V^\dagger_{\rm A/B}
\ ,\label{Eq:A_single}
\end{align}
which can be expressed using Wilson lines,
\begin{align}
V^\dagger_{\rm A/B}
= 
P_{\xmp} 
\exp{\left[ -i \int^{\xmp}_{-\infty} 
dx'^\mp \partial^{-2}_\perp 
\rho_{\rm A/B,cov} \right]}
\ ,\label{Eq:V}
\end{align}
where the indices A and B in the gauge field and Wilson line represent nuclei A and B, respectively, and $P_{\xmp}$ is the $\xmp$-ordered product.
The classical color charge density in the Wilson line, $\rho_{\rm cov}$, is a solution of the continuity equation shown in Eq.~\refeq{Eq:current} with the gauge field in the covariant gauge $\partial_\mu A^\mu=0$, while the gauge condition for the gauge fields in Eq.~\refeq{Eq:A_single} is given in the light-cone gauge, $A^{{\rm A} +}=0$ or $A^{{\rm B} -}=0$.
The classical color charge densities in the covariant and light-cone gauges can be transformed into each other by a gauge transformation,
\begin{align}
\rho_{\rm A/B, cov} 
= V^\dagger_{\rm A/B} 
\rho_{\rm A/B, (L.C.)} V_{\rm A/B}\ .\label{Eq:gt}
\end{align}
An important feature of the classical color charge density in the light-cone gauge is that it is independent of the light-cone time, namely static: $\partial_{\pm} \rho_{\rm A/B}=0$.
Moreover, in this gauge, the CYM fields also become static: $\partial_{\pm} A^{\rm A/B}=0$.
Later, we simply denote the classical color charge density in the covariant gauge as $\rho$ by omitting ``cov": $\rho_{\rm cov} \to \rho$.

Before moving to the case of the collision, let us emphasize that the WW fields given in Eq.~\refeq{Eq:A_single} have only the transverse colorelectric and colormagnetic fields, similar to the WW fields of the electromagnetic field.
The colorelectric and colormagnetic fields in the light-cone coordinates are defined as $E^{{\rm A/B},i} \equiv F^{i \pm}$ and $B^{{\rm A/B},i} \equiv \epsilon^{ijk} F_{jk}\ \ \ (i,j,k=1,2,\mp)$, respectively.
It is easy to obtain the colorelectric and colormagnetic fields of the WW fields as follows:
\begin{align}
E^{{\rm A/B},1} 
&=
B^{{\rm A/B},2} 
=
\frac{1}{g} 
V^{\rm A/B}  
\left[
\frac{\partial_1}{\partial^2_\perp} 
\rho^{\rm A/B}
\right]
V^{{\rm A/B} \dagger}
\ ,\\
E^{{\rm A/B},2} 
&=
-B^{{\rm A/B},1} 
=
\frac{1}{g} 
V^{\rm A/B}  
\left[
\frac{\partial_2}{\partial^2_\perp} 
\rho^{\rm A/B}
\right]
V^{{\rm A/B} \dagger}
\ ,\\
E^{{\rm A/B},\mp} 
&=
B^{{\rm A/B},\mp}=0\ .\label{Eq:EandB}
\end{align}
Next, we consider the case of a collision occurring.
Then, the complexity of the CYM equation of motion and continuity equation increases significantly compared to the case of a single nucleus.
Obtaining an analytical solution without approximations becomes impractical.
Numerically solving the evolution equations is also challenging.
However, the approximations of infinitely thin nuclei and their remaining unchanged due to the collision are known to drastically simplify the treatment, 
although these restricts the CYM equation of motion to being boost invariant~\cite{MVcollision,2Dglasma_KV}.
We try to solve those evolution equations exactly without these approximations in the modified Milne coordinates, starting from the initial conditions before the collision until the time of interest after the collision.
The gauge condition used is the Fock-Schwinger gauge for the modified Milne coordinates, $A_{\ttau}=0$.

\begin{figure*}[tp]
    \centering
    \includegraphics[width=0.5\textwidth]{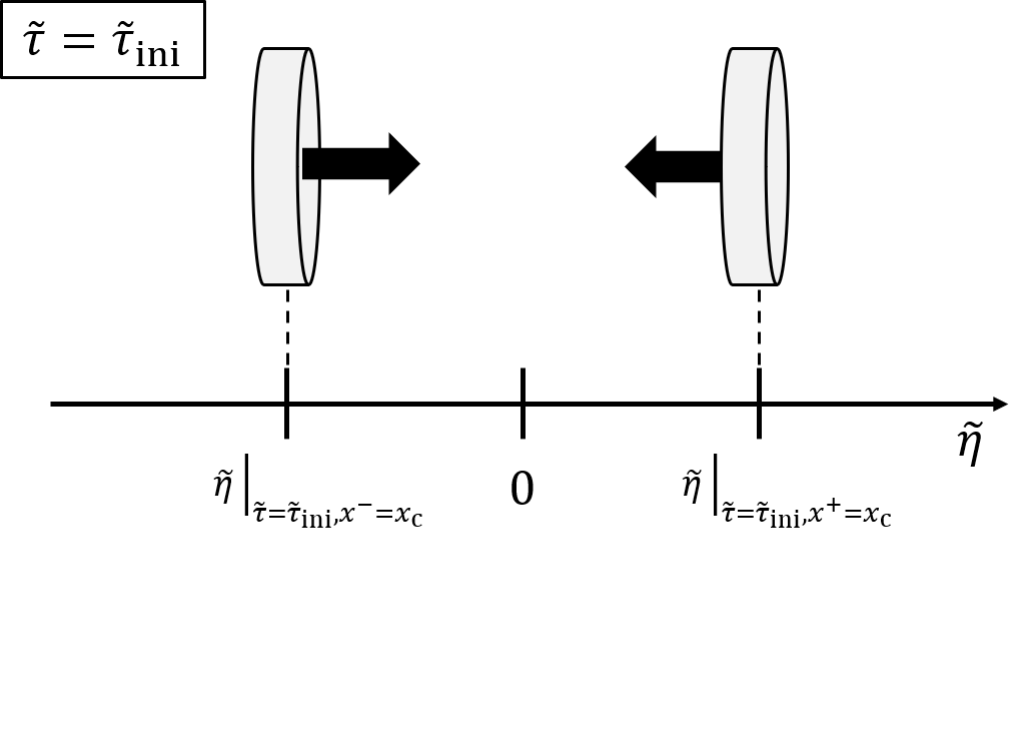}
\caption{
Intuitive depiction of two colliding nuclei at the initial modified proper time $\ttau=\ttau_{\rm ini}$.
}
 \label{Fig:init}
\end{figure*}
The initial modified proper time $\ttau_{\rm ini}$ is chosen to be sufficiently small.
This ensures that the classical color charge densities from nuclei A and B are well separated into the negative and positive modified rapidity regions, respectively, as shown in Fig.~\ref{Fig:init}.
This initial condition corresponds to a situation where the nuclei have not yet collided.
For both nuclei A and B, the single nucleus's solution of the CYM field shown in Eq.~\refeq{Eq:A_single} satisfies the Fock-Schwinger gauge.
Thus, if the two nuclei are completely uncausal at $\ttau = \ttau_{\rm ini}$, the initial conditions for the CYM fields and classical color charge densities are exactly given by the static solutions of the individual nuclei.
However, in practice, we describe the classical color charge densities of the nuclei as the sum of nucleon's color charge densities with a longitudinal Gaussian shape, as explained in the next section.
Therefore, even in the infinite past, there is a small overlap between both nuclei.
As a result, they are always causal for each other in principle.
Thus, as an approximate treatment, we set the initial condition of the CYM fields, except for the longitudinal electric fields, as a superposition of the WW fields from each individual nucleus in the infinite past
\begin{align}
&
A_{i=1,2}|_{\ttau=\ttau_{\rm ini}}       
= 
A^{\rm A}_i + A^{\rm B}_i \ ,\label{Eq:CYMini0}\\
&
A_{\teta}|_{\ttau=\ttau_{\rm ini}}
=
A_{\ttau}|_{\ttau=\ttau_{\rm ini}}
=0\ ,
\end{align}
and
\begin{align}
E^{i=1,2}
|_{\ttau=\ttau_{\rm ini}}
\equiv 
\ttau F^{i \ttau}
|_{\ttau=\ttau_{\rm ini}}
=
\xm E^{{\rm A},i}
+
\xp E^{{\rm B},i}\ .
\end{align}
Here, the general coordinate transformation from the light-cone coordinates to the modified Milne coordinates is used.
In addition to these, the longitudinal colorelectric field, $E^{\teta}\equiv \ttau F^{\teta \ttau}$, is given by
\begin{align}
E^{\teta}
|_{\ttau=\ttau_{\rm ini}}
= 
ig[A^{\rm A}_1, A^{\rm B}_1]
+
ig[A^{\rm A}_2, A^{\rm B}_2]\ ,
\label{Eq:CYMini2}
\end{align}
which is determined to make the initial condition satisfy Gauss's law,
\begin{align}
\sum_{i=1,2,\teta} 
[D_i, E^i]=
\frac{1}{\ttau}
\left[ \xm \rho_{\rm A, (L.C.)} + \xp \rho_{\rm B, (L.C.)} \right]\ .
\end{align}
Here, $\rho_{\rm A/B}$ is given as event-by-event random numbers according to a specified probability distribution, 
and $\rho_{\rm A/B, (L.C.)}$ is obtained from $\rho_{\rm A/B}$ via a gauge transformation.
The longitudinal colorelectric field, given in Eq.~\refeq{Eq:CYMini2}, vanishes if there is no overlap between the two nuclei at the initial modified proper time $\ttau_{\rm ini}$, namely if they are completely uncausal.

The behavior of the system for the later time, $\ttau > \ttau_{\rm ini}$, is obtained by solving the CYM equation of motion and continuity equation with respect to $\ttau$, starting from the aforementioned initial conditions.
The evolution of the system includes the collision and separation of nuclei, as well as the generation and evolution of glasma.
In actual calculations, the initial conditions are discretized on a spatial lattice, and the equations of motion and continuity equation for the discrete fields are numerically solved to track the $\ttau$-evolution of the fields.
The lattice formulation is provided in Ref.~\cite{3Dglasma_FU1}.

\section{Initial condition of nucleus with finitie longitudinal thickness}\label{Sec:3}
Here, we construct a nucleus model for the pre-collision gold nucleus at $\snn = 200$ GeV.
We simply extend the infinitely thin nucleus model used in the boost-invariant IP-glasma model to a nucleus with longitudinal structure, considering the finite thickness of nucleons and their random positions along the collision axis.
We consider the initial condition of the classical color charge density of the nucleus as a superposition of nucleon's color charge densities,
\begin{align}
\rho_{\rm A/B} =
\sum^{N_{\rm A}}_{i=1} \rho_{{\rm A/B},i}\ ,
\end{align}
where the atomic number is taken as gold's value, $N_{\rm A} = 197$.
The positions of the $i$-th nucleon for each nucleus, $(x^1_{{\rm A/B},i},x^2_{{\rm A/B},i},\xmp_{{\rm A/B},i})$, are randomly assigned for each event according to a Woods-Saxon distribution,
\begin{align}
f_{ \rm ws}(\bxp,x^\mp) 
\propto 
\frac{1}
{1+e^{\frac{\sqrt{(\bxp-\bbp)^2+2(x^\mp-b^\mp)^2/\gamma^2}-R}{a_{\rm skin}}}}\ ,
\end{align}
where $\gamma=108$ is the Lorentz factor, $R=6.38$ fm is the nucleus radius, $a_{\rm skin}=0.535$ fm is the thinness of the nucleus surface, and $\bb$ is the center position of the nucleus, which is set to $(b^1,b^2,b^-) = (-\imp/2,0,\xc)$ for nucleus A and $(b^1,b^2,b^+) = (\imp/2,0,\xc)$ for nucleus B.
Here, $\imp$ is the impact parameter.

The initial color charge density of the $i$-th nucleon is given as a random number to satisfy the following event averages:
\begin{align}
\average{ 
\rho^{a}_{{\rm A/B},i}(\bx )
}
= 0\ ,
\end{align}
and
\begin{align}
\average{ 
\rho^{a}_{{\rm A/B},i}(\bx )
\rho^{b}_{{\rm A/B},i}(\bx')
}
&= 
\delta^{a,b} 
\left[ g^2\mu
\left(\frac{\bx+\bx'}{2}-\bx_{{\rm A/B},i}\right)
\right]^2
\nonumber\\
&\times
\no(\xmp - x'^\mp;\cl)
\delta^2(\bxp-\bxp')\ ,\label{Eq:rho_ave1}
\end{align}
where $a$ is a color index, $\mu^2$ denotes the average squared color charge of the hard partons within the nucleon, which is the MV model parameter appearing as $g^2\mu$, and $\no(\xmp-x'^\mp;\cl)$ is a normal distribution function with a variance of $\cl$ that controls the longitudinal correlation length in the color charge density. 
These event-averaged relations are a generalization of the original MV model, where the color charge density is independent for each point in space, which is recovered in the small limit of $\cl$.

The nucleon's color charge density is assumed to have a Gaussian form in the longitudinal direction:
\begin{align}
\rho^a_{{\rm A/B},i}(\bx) 
= 
\no(\xmp-\xmp_{{\rm A/B},i};\sqrt{2}\rl) 
\times \Gamma^a_{{\rm A/B},i}(\bx)\ ,\label{Eq:rho_assum}
\end{align}
where the variance of the normal distribution, $\sqrt{2}\rl$, corresponds to the spatial spread of the nucleon's color charge density, and $\Gamma_{A/B,i}$ is a random number satisfying the following event average,
\begin{align}
&
\average{ 
\Gamma^a_{{\rm A/B},i}(\bx )
\Gamma^b_{{\rm A/B},i}(\bx')
}
= 
\delta^{a,b}
\sqrt{2\pi\left(2r^2_{\rm L}+\sigma^2_{\rm L}\right)}
\nonumber\\
&\times
\left[ g^2\mu_{\rm 2D}(\bxp) \right]^2
\no(\xmp-x'^\mp;\sigl)
\delta^2(\bxp-\bxp')\ ,\label{Eq:gamma_i}
\end{align}
with the longitudinal correlation length $\sigl$, which relates $\cl$ and $\rl$ as
\begin{align}
\sigl^{-2} = \cl^{-2} - (2\rl)^{-2}\ .\label{Eq:sigl}
\end{align}
This relation provides a constraint on $\cl$ as $\cl < 2 \rl$, reflecting the limitation of the longitudinal correlation length by the longitudinal extent of gluons in the nucleon.
It is easily seen that the classical color charge density given in Eq.~\refeq{Eq:rho_assum} satisfies Eq.~\refeq{Eq:rho_ave1} with
\begin{align}
\left[ g^2\mu(\bx) \right]^2
=
\left[ g^2\mu_{\rm 2D}(\bxp) \right]^2
\times
\no(\xmp;\rl)\ ,\label{Eq:ggmu}
\end{align}
where $[g^2\mu_{\rm 2D}]^2 = \int d\xmp [g^2\mu]^2$ is an integration of $[g^2\mu]^2$ over the light-cone space.

Usually, the MV model parameter is considered through the saturation scale, $\qs$, which is the characteristic transverse momentum scale for soft gluons with the Bjorken $x$ of interest inside a nucleus or nucleon.
Especially, the IP-glasma model refers to the phenomenological values of the saturation scale estimated using the IP-sat (Impact-Parameter dependent saturation) model~\cite{IPsat1,IPsat2}.
The saturation scale in the IP-sat model depends not only on Bjorken $x$, but also on the transverse coordinate, reflecting the transverse distribution of soft gluons.
We use the proton's saturation scale $\qsn$ estimated by the IP-sat model with the HERA data~\cite{IPsat_HERA}, and define $g^2 \mu_{\rm 2D}$ as 
\begin{align}
g^2 \mu_{\rm 2D} = \lambda \qsn,
\end{align}
where $\lambda$ is a parameter that controls their ratio.

To determine the saturation scale $\qsn$, we first specify the Bjorken $x$.
Unlike deep inelastic scattering (DIS), the value of the Bjorken $x$ for the incoming partons contributing to observables cannot be specified uniquely by kinematics.
This is because the glasma is a collection of the soft gluons liberated by the collision of two CGCs.
Thus, we need to determine the ``effective" $x$ of soft gluons in the colliding nuclei that is relevant to the glasma produced.
Referring to the on-shell condition for a single parton,
we introduce the following assumption for such an effective $x$~\cite{Lappi}:
\begin{align}
x \sim 
\frac{p^{\rm typ}_\perp}{\snn} e^{Y_{\rm ref}}\ ,\label{Eq:xeff}
\end{align}
where $p^{\rm typ}_\perp$ is a typical transverse momentum, which is set here as $p^{\rm typ}_\perp \sim 1$ GeV, and $Y_{\rm ref}$ is the momentum rapidity of the interested gluon in the glasma.
Since gluons are massless and the collision happens in a narrow region along the collision axis, the momentum rapidity is roughly interchangeable with the (spacetime) rapidity, $Y_{\rm ref} \sim \eta$.
Later, we take the effective $x$ as $x = 0.01$, which is consistent with near-central rapidity $\eta \sim 0$.
Thus, although the current model accounts for rapidity-dependent effects beyond the shock-wave approximation, it can be regarded as being tuned more reliably for the mid-rapidity region.

We then address the model parameters in the Gaussian functions, $\rl$ and $\cl$.
We remind the reader that the classical color charge density corresponds to hard partons with a light-cone momentum larger than that of the soft gluon with $x$, $p^\pm = x P^\pm_{\rm Hadron}$, where $P^\pm_{\rm Hadron} \sim \snn/\sqrt{2}$ is a hadron's light-cone momentum.
Qualitatively, the inverse of the longitudinal momentum $p^\pm$ gives the longitudinal spread of the color charge density in light-cone coordinates, $\Delta x^\mp \sim 1/p^\pm=\sqrt{2}/(x \snn)$.
Thus, we set $\rl$ to satisfy 
\begin{align}
2 \sqrt{2} \rl = \sqrt{2}/(x \snn)=\sqrt{2}/(0.01 \cdot 200)\;\; {\rm GeV}.
\end{align}
The longitudinal correlation length $\sigl$ is a free parameter in the model, and, naively speaking, reducing $\cl$ would account for the short-distance structure of the classical color charge density carried by hard partons with larger $x$.
The influence of this parameter on various physical quantities has been intensively studied using the semi-analytical method~\cite{3Dglasma_analytic1,3Dglasma_analytic2} and numerical simulation~\cite{3Dglasma_FU2}.
Here, we choose the value of $\cl$ ranging from the maximum value $\cl = 2\rl$ to a smaller one by $0.4$, $\cl = 0.4 \cdot 2\rl$, and investigate the impact of the choice of $\cl$ on physical observables within our setup.

\begin{figure*}[tp]
    \centering
    \includegraphics[width=0.25\textwidth]{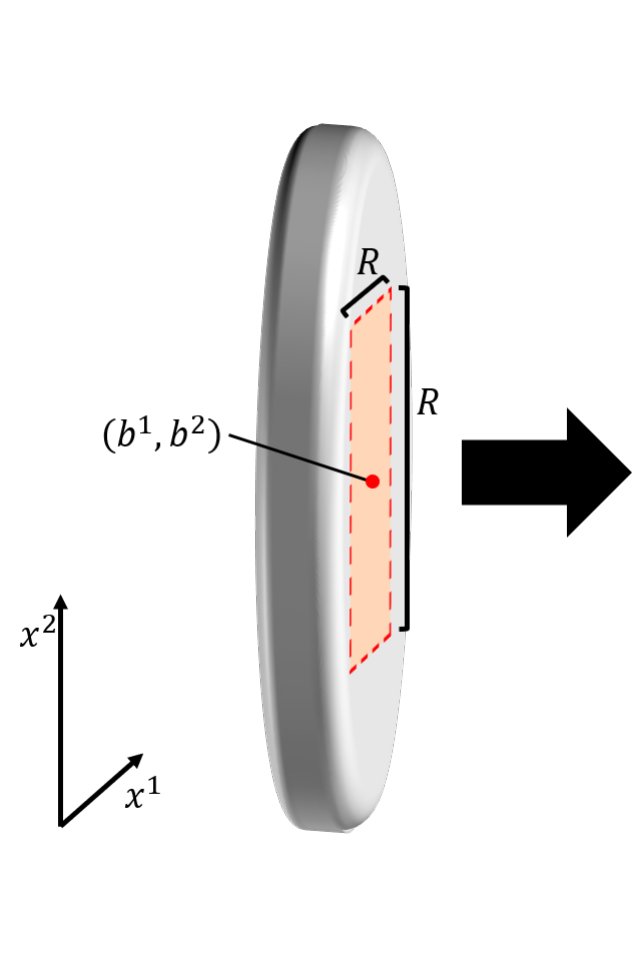}
\caption{
The central region of the nucleus in the transverse plane, 
defined as $\{(x^1, x^2) \in \mathbb{R}^2 \mid -R/2 < x^1 - b^1, x^2 - b^2 < R/2\}$.
}
 \label{Fig:init}
\end{figure*}
We choose the ratio $\lambda$ such that the saturation scale of the nucleus estimated using this nuclear model reproduces the phenomenological estimate of the saturation scale.
This approach to determine the ratio between the MV model parameter and the saturation scale was first proposed in Ref.~\cite{Lappi}.
This approach involves extracting the saturation scale of the nucleus from the peak structure in the Fourier transform of the correlation function of the Wilson line for an infinitely large value of the light-cone space variable:
$V_\infty \equiv \lim_{\xmp\to\infty} V_{\rm A/B}$,
\begin{align}
&
k^2_\perp C(\bkp)
\nonumber\\
&
= 
\int d^2\bxp
\average{\left|{\rm Tr} 
\left[V^\dagger_\infty(\bxp+\bm{y}_\perp) 
V_\infty(\bm{y}_\perp)\right] \right|^2-1} e^{i \bkp \cdot \bxp}\ ,\label{Eq:ft_cw}
\end{align}
where the correlation function does not depend on $\bm{y}_\perp$ due to the translational invariance in the transverse plane.
Since our model does not have exact transverse translational invariance, we only consider the central region of the nucleus, defined as  $\{(x^1, x^2) \in \mathbb{R}^2 \mid [-R/2<x^1-b^1,x^2-b^2<R/2]\}$, 
where we assume approximate translational invariance holds (Fig.\ref{Fig:init}).
Then, we choose the ratio $\lambda$ such that the extracted saturation scale of the nucleus in this central region reproduces the maximal value of the saturation scale of gold, $\qsa=1.1$ GeV, obtained using the IP-sat model~\cite{IPsat_A}.

\begin{figure*}[tp]
\begin{minipage}{0.45\textwidth}
    \centering
    \includegraphics[width=\textwidth]{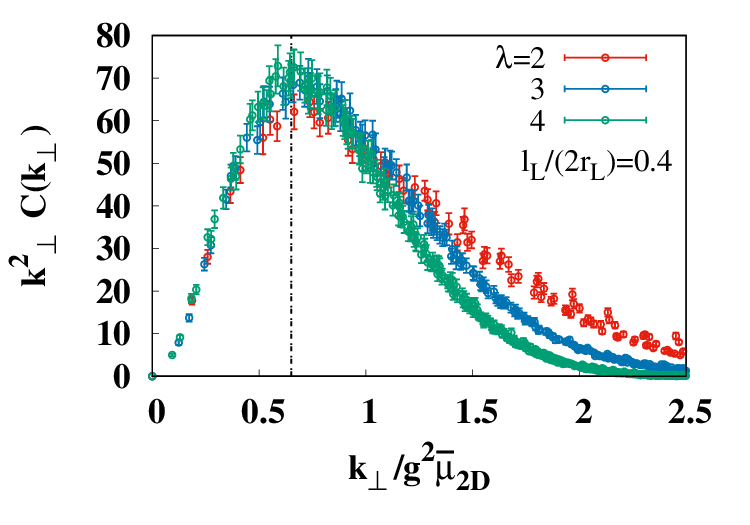}
\end{minipage}%
\hfill
\begin{minipage}{0.45\textwidth}
    \centering
    \includegraphics[width=\textwidth]{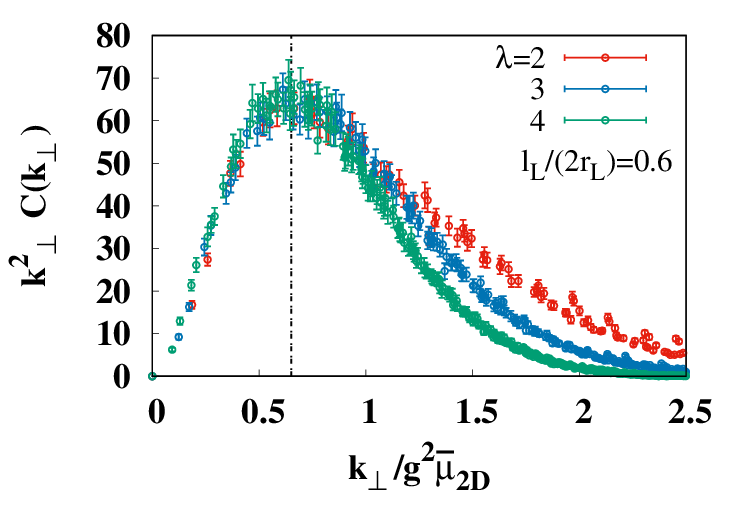}
\end{minipage}
\begin{minipage}{0.45\textwidth}
    \centering
    \includegraphics[width=\textwidth]{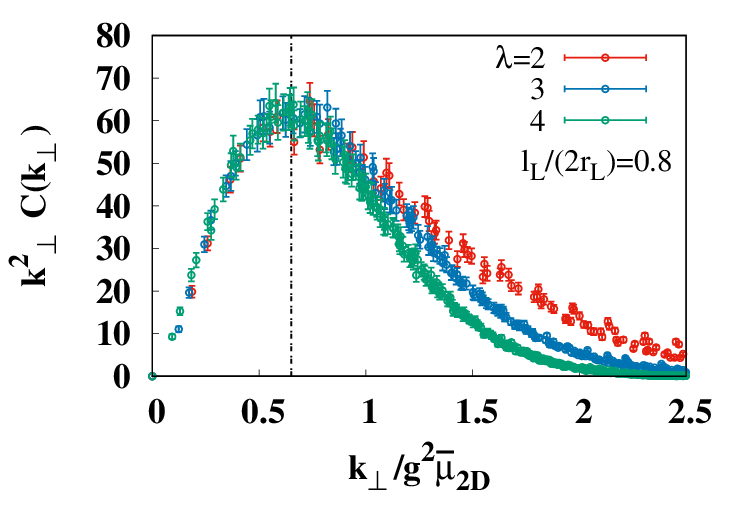}
\end{minipage}%
\hfill
\begin{minipage}{0.45\textwidth}
    \centering
    \includegraphics[width=\textwidth]{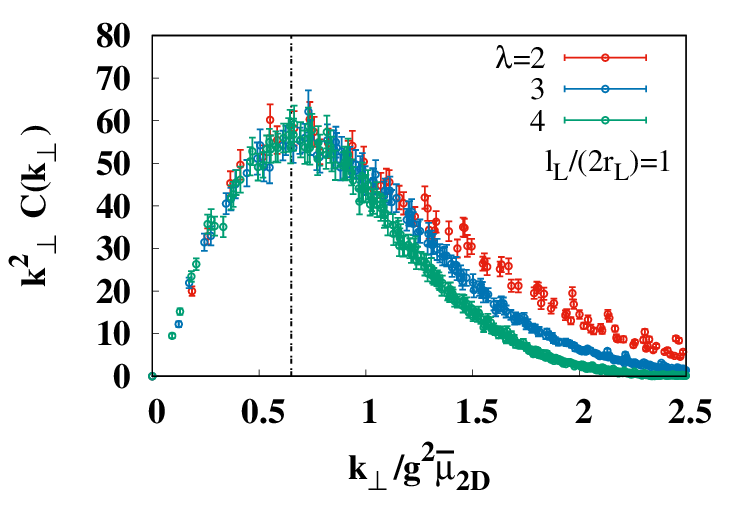}
\end{minipage}
\caption{
The transverse momentum dependence of the Fourier transform of the Wilson line correlation function in the central region.
Shown for longitudinal correlation lengths $\cl/2\rl=0.4,0.6,0.8,1$ and ratios $\lambda=2,3,4$.
The vertical axis is normalized by the squared transverse momentum, and the horizontal axis is normalized by the maximum value of $g^2\mu_{\rm 2D}$.
The top row shows results for $\cl/2\rl=0.4$ and $0.6$. The bottom row shows results for $\cl/2\rl=0.8$ and $1.0$.
The black dashed line denotes $k_\perp/g^2\mu_{\rm 2D}=0.65$.
}
\label{Fig:qs}
\end{figure*}

Figure~\ref{Fig:qs} shows the transverse momentum dependence of the Fourier transform of the Wilson line correlation function in the central region.
This is presented for different values of the longitudinal correlation length, $\cl/2\rl = 0.4, 0.6, 0.8$, and $1$, and the ratios, $\lambda = 2, 3$, and $4$. The vertical axis is normalized by the squared transverse momentum, and the horizontal axis is normalized by the maximum value of $g^2\mu_{\rm 2D} = \lambda \qsn$, $g^2\bar{\mu}_{\rm 2D} \sim 0.532 \lambda$ GeV.
The error magnitude is calculated as the unbiased variance divided by the square root of the event count.
Here, the infrared (IR) and ultraviolet (UV) regulators, explained below, are set to $\ir = 0.2$ GeV and $\uv = 2$ GeV, respectively.
Interestingly, for all values of $\lambda$ and $\cl$, the peak appears around $k_\perp/g^2\bar{\mu}_{\rm 2D} \sim 0.65$.
By using this relation, we can obtain the saturation scale in the central region as $Q^{\rm estimated}_{\rm s,A} = 0.65 g^2\bar{\mu}_{\rm 2D} \sim 0.346 \lambda$ GeV for $2 \leq \lambda \leq 4$ and $0.4 \leq \cl/2 \leq 1$. 
Thus, the appropriate $\lambda$ can be fixed to reproduce the phenomenological value of the nucleus's saturation scale as $\lambda = 1.1[{\rm GeV}]/0.346[{\rm GeV}]\sim3.18$.

It should be mentioned that, although the ratio $k_\perp/g^2\bar{\mu}_{\rm 2D}$ for the peak position doesn't change within the range of parameters, $2 \leq \lambda \leq 4$ and $0.4 \leq \cl/2 \leq 1$, generally it can depend on those parameters.
In fact, it is found in Ref.~\cite{Lappi} that their results depend on the ratio of $g^2\mu$, which corresponds to $g^2\mu_{\rm 2D}$ in our model, and the IR regulator $\ir$, as well as the degree of longitudinal randomness in the classical color charge density, which is accounted for by changing $\cl$ in our model.

Note that some simplifications are imposed here, unlike the conventional 2D IP-glasma model.
The first is that we give each nucleon's color charge density based on the saturation scale from the IP-sat model, while normally one utilizes the nucleus's saturation scale to obtain the classical color charge density of the nucleus directly.
Instead of this simplification, we can account for the event-by-event longitudinal position of nucleons in the nucleus model.
Accompanying the above simplification, we further simplify by estimating $x = 0.01$ using the assumed relation given in Eq.~\refeq{Eq:xeff}, while normally one replaces $p^{\rm typ}_\perp$ with the saturation scale $Q_{\rm s}(x)$ and solves it self-consistently to obtain $x$.

In the final part of this section, we explain the IR and UV regulators, $\ir$ and $\uv$, introduced for the classical color charge density by replacing the classical color charge density obtained in the above nucleus model as follows:
\begin{align}
\rho(\bk) \to  
\theta(k_\perp-\uv)
\frac{k^2_\perp}{k^2_\perp + \ir^2} 
\rho(\bk)\ ,
\end{align}
where $k_\perp$ is the absolute value of the transverse momentum.
First, let us discuss the IR regulator.
The Fourier component of the color charge density with transverse momentum $k_\perp$ corresponds to the color charges in hard partons seen by soft gluons with a transverse resolution scale of $1/k_\perp$.
When $1/k_\perp$ becomes larger than the size of the nucleon, the total color charges contained within this seen transverse area become zero due to color neutrality.
Therefore, such IR modes should vanish, and it is necessary to introduce an IR regulator to remove these IR contributions.
We set the IR regulator to be of the order of the QCD scale, $\ir=0.2$ GeV, whose inverse is roughly consistent with the size of a nucleon.
Next, we explain the UV regulator.
As the transverse momentum increases, the transverse area seen by soft gluons becomes very narrow, and eventually, the number of hard partons within the seen transverse area becomes small.
The CGC treatment relies on the assumption that the number of hard partons seen by soft gluons is sufficiently large, which allows hard partons to be modeled as random and classical color charge density.
Thus, too large transverse momentum modes of the classical color charge density are inconsistent with the CGC treatment.
To reflect this limitations of the model, it is necessary to introduce a UV regulator.
However, given that the UV regulator is taken to be sufficiently large, it has been confirmed that the late-time behavior of the glasma is insensitive to the UV cutoff~\cite{SU2vsSU3,FG}.
We set the UV regulator to be $\uv = 2$ GeV.

\section{Physical Quantities for Central Collisions}\label{Sec:4}
In this section, we present the results of simulations for central collisions, conducted with an impact parameter of zero.
First, we calculate the rapidity profile of the energy density in the local rest frame (LRF) for various values of the correlation length $\cl$.
We show the dependence of their shapes on $\cl$ and their magnitudes at mid-rapidity.
Next, we show the rapidity profiles of the energy density and pressure, which are basic quantities of hydrodynamics.
Finally, we investigate the behavior of topological charge and axial charge.

Simulations in this section and the next are performed on a spatial lattice with the number of transverse and longitudinal grid points being $224$ and $896$, respectively, in the modified Milne coordinates $(\ttau,\teta)$.
The transverse and longitudinal lattice spacings are taken as $a_{\rm T}=0.8$ GeV$^{-1}$ and $a_{\teta}=0.005$.
The lattice size and spacing are chosen to ensure that the numerical results are robust against changes in them.
The evolution of the system is obtained by solving the discretized Hamilton equation of motion and continuity equation for the discretized CYM field and classical color charge densities on the lattice using the leap-frog method.
The results are shown in the usual Milne coordinates $(\tau,\eta)$.
To obtain the EM tensor and field strength in the usual Milne coordinates, we perform a general coordinate transformation.
The maximum value of the usual proper time $\tau$ sampled for observables is $\tau=0.6$ fm/c, the phenomenological proper time for the onset of hydrodynamic evolution.
Unless otherwise noted, the observables presented are examined by taking their event average with a number of events being $56$.
The error is estimated in the same manner as for the numerical results of the Wilson line correlator in the previous section.
The number of colors is set to $2$, and the coupling constant is taken as $g = 2$.
It is important to note that, in the CYM calculations, the $g$ dependence of the fields is trivial, which is the pre-factor, as $A, E, J \propto 1/g$.

\subsection{Energy Density in Local Rest Frame}\label{Sec:Elrf}
We first consider the energy density in the local rest frame, $\elrf$, which we call the LRF energy density, defined as an eigenvalue corresponding to the time-like eigenvector of the EM tensor, $T^\mu_\nu u^\nu = \elrf u^\mu$.
The LRF energy density is one of the basic quantities of relativistic hydrodynamics.
We etract the local energy density using the following relation~\cite{3Dglasma_SS}:
\begin{align}
\elrf =
\frac{1}{2}
\left(
\left[
T^{\tau\tau}
-
T^{\eta\eta}
\right]
+
\sqrt{
\left[
T^{\tau\tau}
+
T^{\eta\eta}
\right]^2
-
4[T^{\tau\eta}]^2
}
\right)
\ , \label{Eq:Ene_lrf}
\end{align}
Here, we asuume that the transverse flow is neglected ($u^1 = u^2 = 0$).
This is because, in the early stage, the system created is still expanding almost longitudinally.
The transverse flow would develop in the later stage due to factors such as the transverse expansion of the hydrodynamic fluid caused by a pressure gradient.
All subsequent calculations of the LRF energy density in this paper are based on this simplification.
To estimate the EM tensor using lattice simulations, we employ the expression of the EM tensor that agrees with the EM tensor in continuous space~\cite{3Dglasma_FU1}.

It should be noted that the EM tensor of the WW fields, $T^{\mu\nu}_{\rm WW}$, strictly yields zero LRF energy density as a result of the relations $T^{\tau\tau}_{\rm WW} - T^{\eta\eta}_{\rm WW} = 0$ and $T^{\tau\tau}_{\rm WW} + T^{\eta\eta}_{\rm WW} = 2\left|T^{\tau\eta}_{\rm WW}\right|$.
Mathematically, $T^{\mu\nu}_{\rm WW}$ is a non-normal matrix and has no eigenvalues.
Physically, the zero value of the LRF energy density reflects that the WW field runs at the speed of light, and thus its local rest frame cannot be defined.
Thus, the LRF energy density would include minimal contributions from the WW field. 
However, in general, the CYM field encompasses all the soft gluons from both the glasma and WW fields, and their contributions cannot be clearly separated.

\begin{figure*}[tp]
    \centering
    \includegraphics[width=0.5\textwidth]{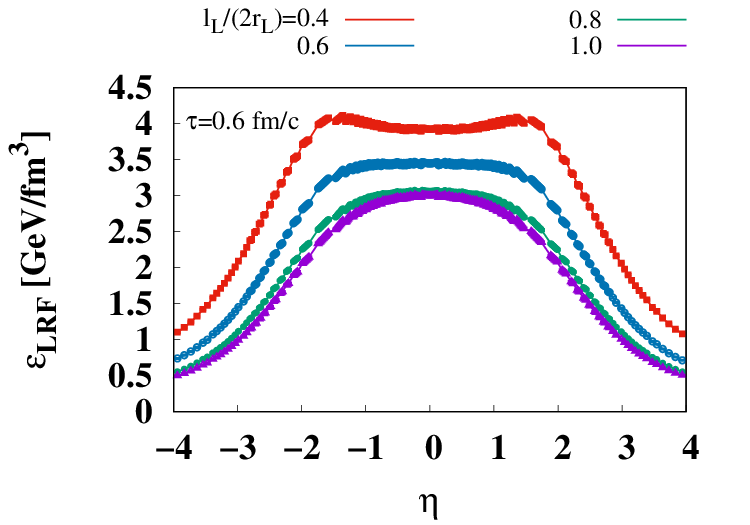}
\caption{
The rapidity profile of the LRF energy density 
$\elrf$, averaged over the transverse plane
in central collisions. 
This is presented for different values of the longitudinal correlation length, 
$\cl/2\rl = 0.4, 0.6, 0.8$, and $1$, at $\tau = 0.6$ (fm/c).
}
 \label{Fig:Ene_lrf}
\end{figure*}
In Fig.~\ref{Fig:Ene_lrf}, we show the rapidity profile of the LRF energy density $\elrf$, integrated over the transverse plane and normalized by $\pi R^2$, 
which approximates the transverse system size of the glasma for central collisions.
Later, we simply refer to this averaging as "averaging over the transverse plane".
The result is presented for different values of the longitudinal correlation length, $\cl/2\rl = 0.4, 0.6, 0.8$, and $1$, at $\tau = 0.6$ fm/c.
It is found that for the maximum correlation length, the shape of $\elrf$ appears more rounded, and as $\cl$ decreases, the plateau structure emerges, develops gradually, and changes into a dip structure.
We here discuss a realistic shape of $\elrf$, assuming there is a close relationship between $\elrf$ and observed hadron multiplicity.
In the RHIC experiment, we have experimental data on the particle-rapidity profile of observed hadrons for central Au-Au collisions at $\snn=200$ GeV: the pion ``rapidity" profile, where the ``rapidity" is different from the spacetime rapidity $\eta$ referred to as rapidity in this paper, shows a Gaussian shape~\cite{RHIC_pion}, and the pseudorapidity distribution of total charged particles shows a plateau (or very slight dip) structure~\cite{RHIC_charged_hadron}.
Even though these particle-rapidities and rapidity $\eta$ may be similar concepts, there is no exact relation between them.
Therefore, $\cl = 0.6$ and $1$, which provide the rounded shape and plateau shape in the (spacetime) rapidity profile of $\elrf$, as shown in Fig.~\ref{Fig:Ene_lrf}, are used for the following simulations.

\begin{figure*}[tp]
\begin{minipage}{0.45\textwidth}
    \centering
    \includegraphics[width=\textwidth]{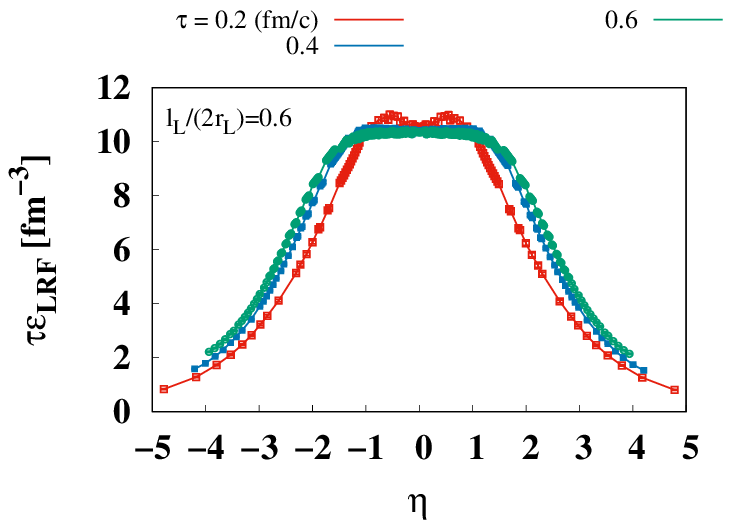}
\end{minipage}%
\hfill
\begin{minipage}{0.45\textwidth}
    \centering
    \includegraphics[width=\textwidth]{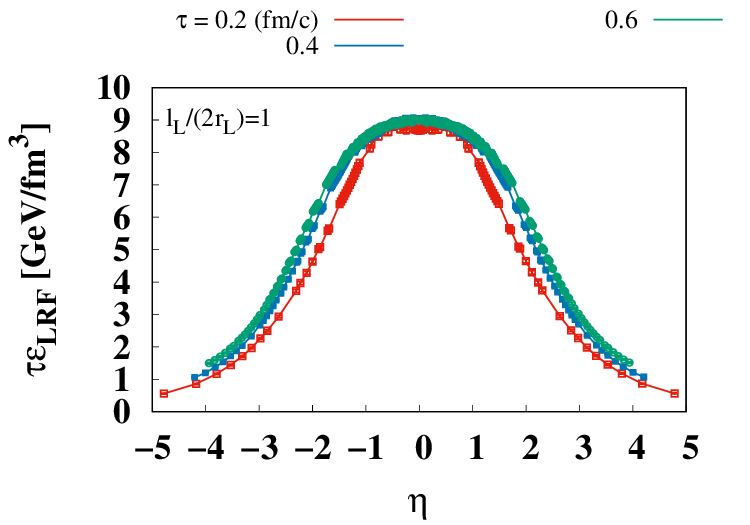}
\end{minipage}
\caption{
The proper time evolution of the rapidity profile of the LRF energy density 
multiplied by the proper time, $\tau\elrf$, averaged over the transverse plane
in central collisions. 
This is presented for $\cl/2\rl=0.6$ (left panel) and $1$ (right panel) 
at $\tau = 0.2, 0.4$, and $0.6$ fm/c.
}
\label{Fig:Ene_lrf_12}
\end{figure*}
In Fig.~\ref{Fig:Ene_lrf_12}, we present the proper time evolution of the rapidity profile of the LRF energy density $\elrf$ 
multiplied by the proper time $\tau$, averaged over the transverse plane in central collisions, 
for different values of the longitudinal correlation length, $\cl/2\rl = 0.6$ and $1$.
It is found that, in the early proper time region, $0.2$ fm/c $< \tau < 0.4$ fm/c, 
the LRF energy density $\elrf$ expands longitudinally faster than $1/\tau$, while in the later proper time region, 
$0.4$ fm/c $< \tau < 0.6$ fm/c, $\elrf$ expands longitudinally at a rate close to being proportional to $1/\tau$.
Especially around mid-rapidity, $\tau\elrf$ appears constant in $\tau$ in the later proper time region.
Here, this $1/\tau$ expansion is consistent with Bjorken flow with no longitudinal pressure.
Therefore, the above finding indicates that the glasma may eventually behave like Bjorken flow without longitudinal pressure after the evolution, and this tendency is more realized near mid-rapidity.

\begin{figure*}[tp]
\begin{minipage}{0.45\textwidth}
    \centering
    \includegraphics[width=\textwidth]{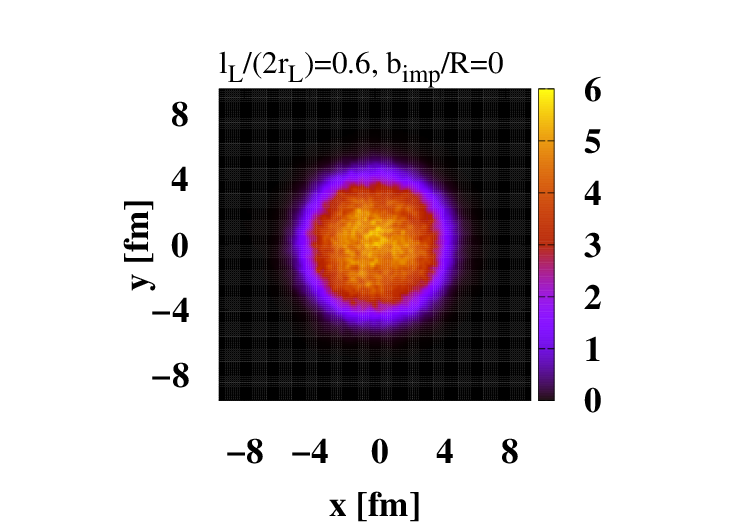}
\end{minipage}%
\hfill
\begin{minipage}{0.45\textwidth}
    \centering
    \includegraphics[width=\textwidth]{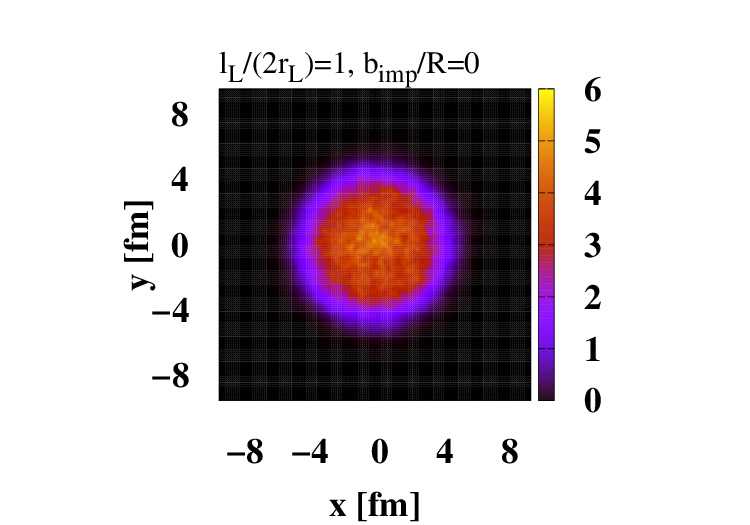}
\end{minipage}
\caption{
The transverse profile of the LRF energy density $\elrf$ 
at mid-rapidity $\eta=0$ in central collisions.
This is presented for $\cl/2\rl=0.6$ (left panel) and $1$ (right panel) 
at $\tau=0.6$ fm/c.
}
 \label{Fig:Ene_lrf_xy}
\end{figure*}
In Fig.~\ref{Fig:Ene_lrf_xy}, we show the transverse profile of the LRF energy density $\elrf$ 
at mid-rapidity $\eta=0$ in central collisions.
This is presented for different values of the longitudinal correlation length, $\cl/2\rl=0.6$ and $1$ 
at $\tau=0.6$ fm/c.
The averaged LRF energy density in the center of the glasma at mid-rapidity, $\bar{\varepsilon} \sim 6$ GeV/fm$^3$, 
is a little smaller than that estimated by boost-invariant hydrodynamic simulations, $\bar{\varepsilon} \sim 11$ GeV/fm$^3$~\cite{Hydro_for_HIC}, 
and the range of the initial energy density estimated by the boost-invariant glasma simulation with MV model, 
$7.1$ GeV/fm$^3\leq\bar{\varepsilon}\leq40$ GeV/fm$^3$~\cite{KNV_energy}.
However, this discrepancy should be improved when changing the number of colors to $N_{\rm c} = 3$.

\subsection{Pressure and Energy Density}\label{Sec:EnePre}
We consider the energy density $\varepsilon$, and the transverse and longitudinal pressures, $P_\perp$ and $P_\eta$, defined as
\begin{align}
\varepsilon &\equiv T^{\tau\tau}\ ,\\
P_\perp   &\equiv \frac{T^{11}+T^{22} }{2}
\ ,\\
P_\eta &\equiv \tau^2 T^{\eta\eta}\ .
\end{align}
The energy density and pressure have the relation $\varepsilon = 2P_\perp + P_\eta$, which is a result of the traceless nature of the EM tensor led by the conformal symmetry of the CYM theory.

\begin{figure*}[tp]
\begin{minipage}{0.45\textwidth}
    \centering
    \includegraphics[width=\textwidth]{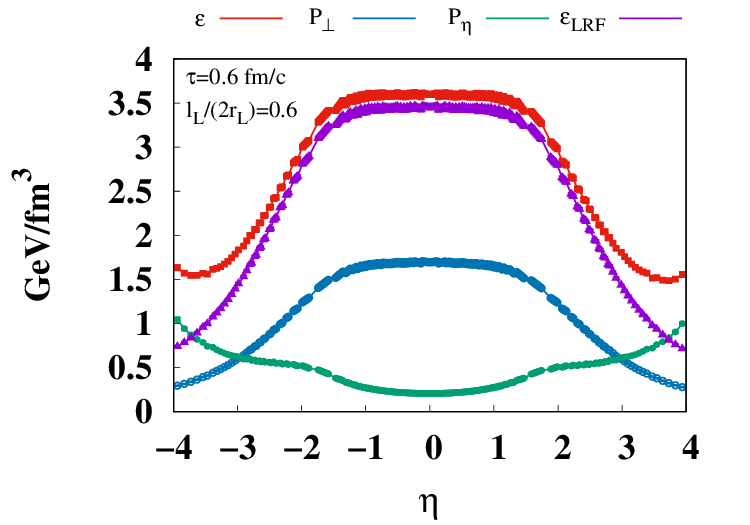}
\end{minipage}%
\hfill
\begin{minipage}{0.45\textwidth}
    \centering
    \includegraphics[width=\textwidth]{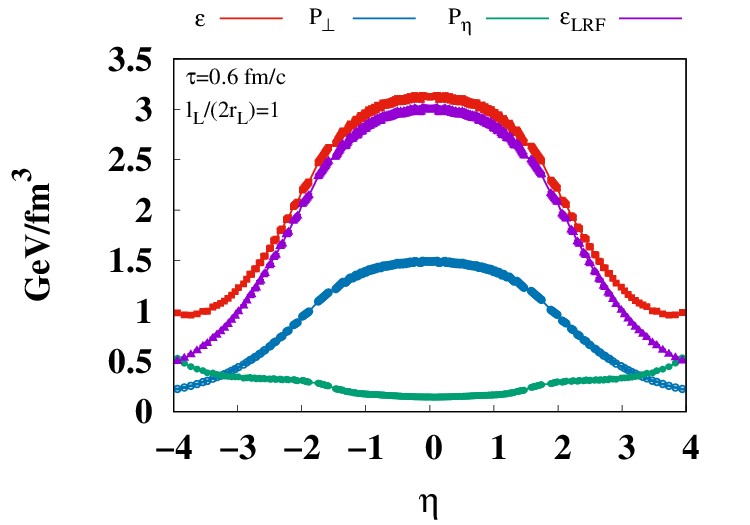}
\end{minipage}
\caption{
The rapidity profile of the energy density $\varepsilon$, 
transverse and longitudinal pressure $P_\perp$ and $P_\eta$, 
and LRF energy density $\elrf$, averaged over the transverse plane in central collisions. 
This is presented for $\cl/2\rl=0.6$ (left panel) 
and $1$ (right panel) at $\tau=0.6$ fm/c.
}
 \label{Fig:EP}
\end{figure*}
In Fig.~\ref{Fig:EP}, we show the rapidity profile of the energy density $\varepsilon$, transverse and longitudinal pressure $P_\perp$ and $P_\eta$, and LRF energy density $\elrf$, 
averaged over the transverse plane in central collisions.
This is presented for different values of the longitudinal correlation length, $\cl/2\rl=0.6$ and $1$ at $\tau=0.6$ fm/c.
Interestingly, while the transverse pressure and LRF energy density are localized near central rapidity, the longitudinal pressure and energy density extend to larger rapidity regions and even increase in larger rapidity regions, as observed also in Refs.~\cite{3Dglasma_FU1,3Dglasma_CPIC,3Dglasma_SS,3Dglasma_FU2,3Dglasma_analytic2}.
Similarly, as shown later, $T^{\tau 3}$ also shows this increase feature even in larger rapidity ranges than those shown in Fig.~\ref{Fig:EP}.
We will revisit and discuss this increase later.

The difference between the energy density and the LRF energy density is worth mentioning. 
This difference is not very large around mid-rapidity but increases significantly at larger rapidity. 
Since this discrepancy stems from the EM tensor not being diagonal, 
it indicates that Bjorken flow, where $u^\tau = 1$ and $u^\eta = 0$, 
no longer accurately describes the longitudinal expansion of the system. 
In fact, in Ref.~\cite{3Dglasma_analytic2}, it is pointed out that the flow of the dilute 3D glasma is slower than the Bjorken flow, especially in the large rapidity region, and this deviation can be explained as if the flow of the 3D glasma is a superposition of the Bjorken flows generated in the collision region.

\subsection{Topological Charge}\label{Sec:CorTop}
The glasma initially has strong parallel longitudinal colorelectric and colormagnetic fields with a transverse correlation length of $1/\qs$ ~\cite{MVcollision,LM}.
This leads to large fluctuations in the topological charge density which is defined as
\begin{align}
n_{\rm T} \equiv \frac{1}{8\pi^2} {\rm Tr} \bm{E} \cdot \bm{B}
=\frac{1}{16\pi^2} 
\left[
\frac{1}{\tau^2}
\left(
  E^1 B^1 + E^2 B^2
\right)
+
  E^\eta B^\eta
\right]\ ,
\end{align}
while the WW field has exactly zero topological charge density since the colorelectric and colormagnetic fields of the WW field are perpendicular to each other, as shown in Eq.~\refeq{Eq:EandB}.
The Adler-Bell-Jackiw anomaly equation tells us that the topological charge density is related to the divergence of the axial current~\cite{ABJ1,ABJ2},
\begin{align}
\partial_\tau (\tau j^\tau_5)
-
\tau \sum_{i=1,2,\eta} \partial_i j^i_5
=
\tau n_{\rm T}\ .
\end{align}
The presence of the axial charge density, $n_5 \equiv \tau j^\tau_5$, may lead to a number of intriguing anomalous transport phenomena such as the Chiral Magnetic Effect. 
Therefore, a study on the topological charges and axial charge density is important for the understanding of these physical quantities in experiments~\cite{CME1,CME2}.
We note that there are already some studies on the topological charge in the boost-invariant glasma~\cite{KKVtopological,SphaleronTransition,Ruggieri_topological}. 
We investigate the rapidity profile of the topological charge and transverse correlation of the axial charge.

\begin{figure*}[tp]
\begin{minipage}{0.45\textwidth}
    \centering
    \includegraphics[width=\textwidth]{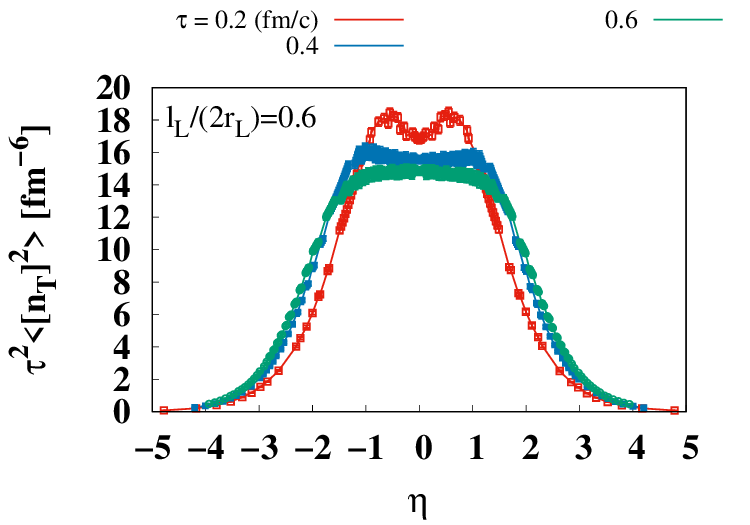}
\end{minipage}%
\hfill
\begin{minipage}{0.45\textwidth}
    \centering
    \includegraphics[width=\textwidth]{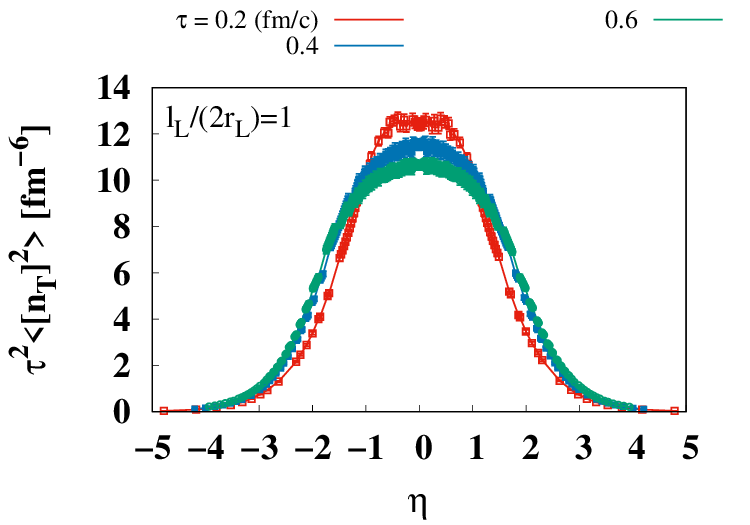}
\end{minipage}
\caption{
The rapidity profile of the square of the product of 
the topological charge density and proper time, 
$\average{(\tau n_{\rm T})^2}$, averaged over the transverse plane
in central collisions.
This is presented for $\cl/2\rl=0.6$ (left panel) and $1$ (right panel) at $\tau=0.6$ fm/c.
}
\label{Fig:nt}
\end{figure*}
In Fig.~\ref{Fig:nt}, we show the rapidity profile of the square of the product of the topological charge density and proper time, $\average{(\tau n_{\rm T})^2}$, 
averaged over the transverse plane in central collisions.
This is presented for different values of the longitudinal correlation length, $\cl/2\rl=0.6$ and $1$, at $\tau=0.6$ fm/c.
These results confirm that topological charge fluctuations are generated around mid-rapidity by the collision, with a shape similar to the LRF energy density. This shows that for central Au-Au collisions at $\sqrt{s_{\rm NN}}=200$ GeV, the topological charge fluctuation averaged over the transverse plane is about $\sqrt{\langle n_{\rm T}^2\rangle}\approx 3-4\; {\rm fm}^{-3}/\tau$ at mid-rapidity.


Here, we consider the transverse correlation function of the axial charge density, $n_5$. We focus on the central area of the glasma in the transverse plane, defined by $-R/2<x^1,x^2<R/2$, 
and at mid-rapidity $\eta=0$, where approximate transverse and longitudinal translational invariance approximately hold.
To estimate $n_5$, we assume that the spatial components of the axial current vanish, $\average{j^{i=1,2,\eta}} = 0$, which is a good approximation due to the approximate translational invariance.
Then, we employ the following relation:
\begin{align}
n_5 = 
\int^\tau
d\tau
\ \tau  n_{\rm T}
=
\int^{\ttau=\ttau(\tau,\eta=0)}
d\ttau\ \tau n_{\rm T}\ ,
\end{align}
where we use $d\tau = d\ttau$ which holds at the mid-rapidity.
The integration area in the modified proper time is bounded such that 
the usual proper time, which is a function of the modified Milne coordinate variables $\tau = \tau(\ttau, \teta)$, 
does not exceed the usual proper time of interest.

\begin{figure*}[tp]
\begin{minipage}{0.45\textwidth}
    \centering
    \includegraphics[width=\textwidth]{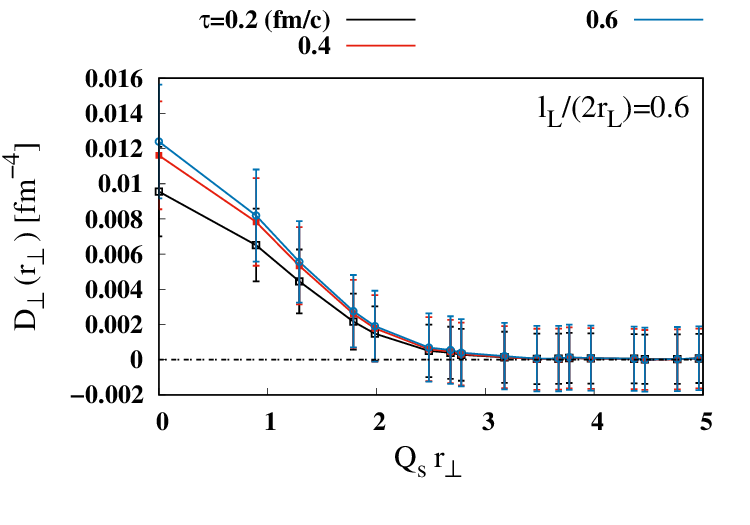}
\end{minipage}%
\hfill
\begin{minipage}{0.45\textwidth}
    \centering
    \includegraphics[width=\textwidth]{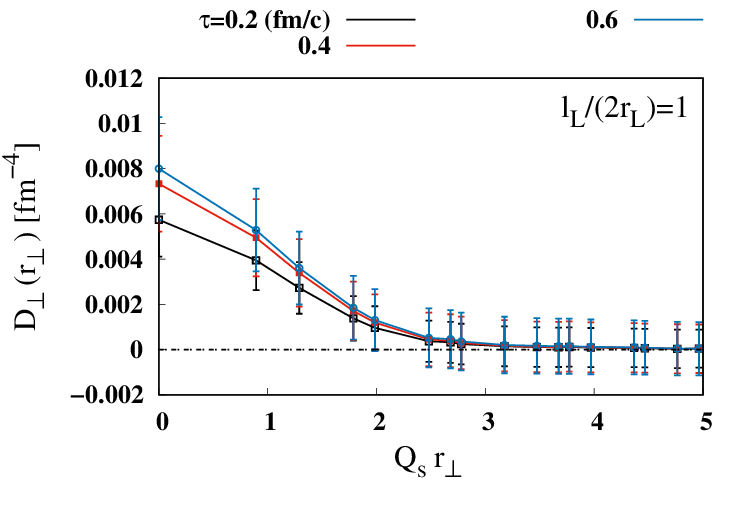}
\end{minipage}
\caption{
The transverse correlation function of 
$n_5$, as defined in Eq.~\refeq{Eq:col}.
This is presented at $\tau=0.2, 0.4$, and $0.6$ fm/c for 
$\cl/2\rl=0.6$ (left panel) and $1$ (right panel). 
Here, the saturation scale is taken as $\qs=1.1$ GeV.
}
\label{Fig:n5_t}
\end{figure*}
In Fig.~\ref{Fig:n5_t}, we show the transverse correlation function 
of $n_5$ defined as
\begin{align}
D_\perp(r_\perp=|\bm{x}_\perp-\bm{y}_\perp|) \equiv 
\average{n_5(x^1,x^2) 
         n_5(y^1,y^2) }\ .\label{Eq:col}
\end{align}
This is presented at $\tau=0.2, 0.4$, and $0.6$ fm/c for 
different values of the longitudinal correlation length, $\cl/2\rl=0.6$ and $1$. The transverse correlation function decreases in the region $\qs r_\perp < 3$, and for $\qs r_\perp > 3$, 
the values are consistent with zero within the error. 
This behavior of the transverse correlation function does not change over proper time evolution. 
These results are in qualitative agreement with previous studies on the boost-invariant glasma~\cite{Ruggieri_topological}, 
considering that the behavior of our results at $\qs=1.1$ GeV falls between the results at $\qs=1$ and $1.7$ GeV presented in that paper. 
The magnitude of 
$\sqrt{\average{n^2_5}}/\tau$ at $\tau=1/\qs$ is approximately $0.5 \text{ fm}^{-3}$ for $\cl/2\rl=0.6$ and 
$0.38 \text{ fm}^{-3}$ for $\cl/2\rl=1$. 
These magnitudes can also be explained as values between the previously estimated values in the boost-invariant glasma 
at $\qs=1$ and $1.7$ GeV~\cite{Ruggieri_topological}.

\section{Impact Parameter Dependence of Physical Quantities}\label{Sec:5}
In this section, we present and discuss the results of calculations for various impact parameters, focusing on impact parameter-dependent quantities.
First, we examine the eccentricity, which is a measure of the spatial anisotropy of the system in the transverse plane perpendicular to the collision axis.
The eccentricity is converted into anisotropic flow during hydrodynamic evolutions.
Next, we investigate the generation of angular momentum caused by the asymmetry of the system's geometry with respect to the impact parameter direction.
This angular momentum, which is perpendicular to the reaction plane, is expected to be converted into the polarization of observed hadrons, called global polarization~\cite{GlobalPolarization1,GlobalPolarization2}.
Finally, we present the vortex perpendicular to the reaction plane to study the rotation of the glasma.

\subsection{Eccentricity}\label{Sec:Eccen}

\begin{figure*}[tp]
\begin{minipage}{0.45\textwidth}
    \centering
    \includegraphics[width=\textwidth]{ExyAve_1_b0_v2.eps}
\end{minipage}%
\hfill
\begin{minipage}{0.45\textwidth}
    \centering
    \includegraphics[width=\textwidth]{ExyAve_2_b0_v2.eps}
\end{minipage}
\begin{minipage}{0.45\textwidth}
    \centering
    \includegraphics[width=\textwidth]{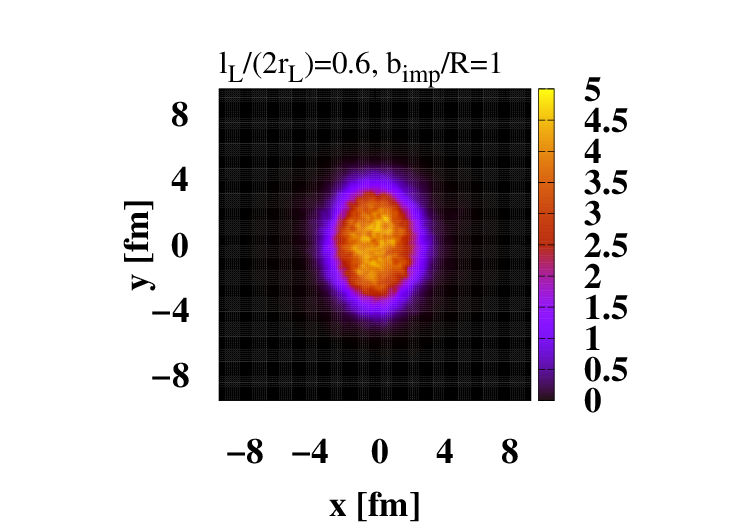}
\end{minipage}%
\hfill
\begin{minipage}{0.45\textwidth}
    \centering
    \includegraphics[width=\textwidth]{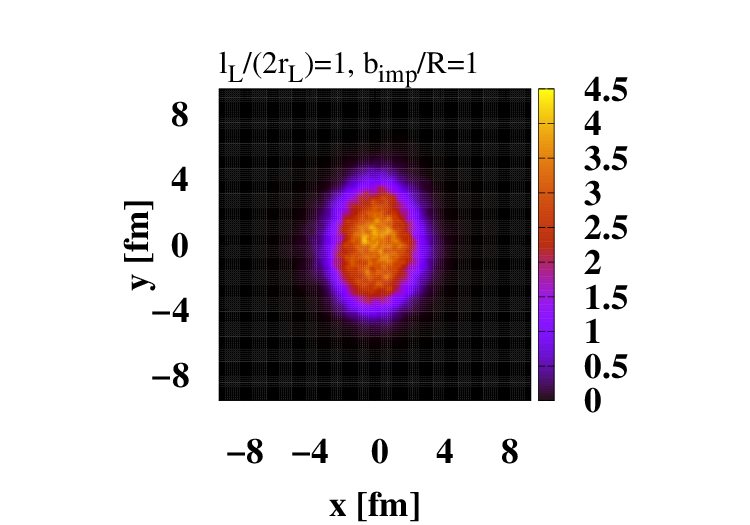}
\end{minipage}
\caption{
The LRF energy density $\elrf$ at mid-rapidity $\eta=0$ 
for two choices of the longitudinal correlation length, 
$\cl/2\rl=0.6$ and $1$, and impact parameter, 
$\imp/R=0$ and $1$ at $\tau=0.6$ fm/c. 
The upper panels show the results for the central collision 
with $\imp/R=0$, 
with the left and right panels showing 
$\cl/2\rl=0.6$ and $1$, respectively. 
The lower panels show the results for the non-central collision 
with $\imp/R=1$, 
with the left and right panels showing $\cl/2\rl=0.6$ and $1$, 
respectively. 
}
\label{Fig:el_tl}
\end{figure*}
In Fig.~\ref{Fig:el_tl}, we show 
the LRF energy density $\elrf$ at mid-rapidity $\eta=0$ 
for two choices of the longitudinal correlation length, 
$\cl/2\rl=0.6$ and $1$, and impact parameter, 
$\imp/R=0$ and $1$ at $\tau=0.6$ fm/c. 
It is found that the LRF energy densities at the central collision, namely for $\imp/R=0$, 
are isotropic in the transverse plane, while those at the non-central collision with $\imp/R=1$ 
have an anisotropic shape that looks elliptical.
Such a deformation of the $\elrf$ shape is expected to reflect 
the shape of the overlap region of the colliding nuclei.

Next, we calculate the eccentricity, a physical quantity characterizing the spatial anisotropy of the produced matter in the transverse plane perpendicular to the collision axis.
We define the eccentricity using the LRF energy density as follows:
\begin{align}
\varepsilon_n 
= 
\frac{
\int d^2\bxp \elrf
r^n_\perp e^{i n \phi}
}{
\int d^2\bxp \elrf r^n_\perp
}\ ,\;\;\;\; n>1\ , \label{Eq:e_n}
\end{align}
where $r_\perp\equiv\sqrt{(x^1)^2+(x^2)^2}$ and $\phi \equiv \arctan{(x^2/x^1)}$ are the polar coordinates in the transverse plane.
This definition of the eccentricity, shown in Eq.~\refeq{Eq:e_n}, is chosen such that the real part of the eccentricity at $n = 2$, which characterizes the elliptical deformation of the glasma, is given by
\begin{align}
{\rm Re} \varepsilon_2 
= 
\frac{
\int d^2\bxp \elrf
[(x^1)^2-(x^2)^2]
}
{
\int d^2\bxp \elrf
[(x^1)^2+(x^2)^2]
}\ .
\end{align}
The eccentricity is converted into anisotropic flow (such as the elliptic flow which is a response to $\varepsilon_2$) during the system's collective evolution in the hydrodynamic stage.

\begin{figure*}[tp]
\begin{minipage}{0.45\textwidth}
    \centering
    \includegraphics[width=\textwidth]{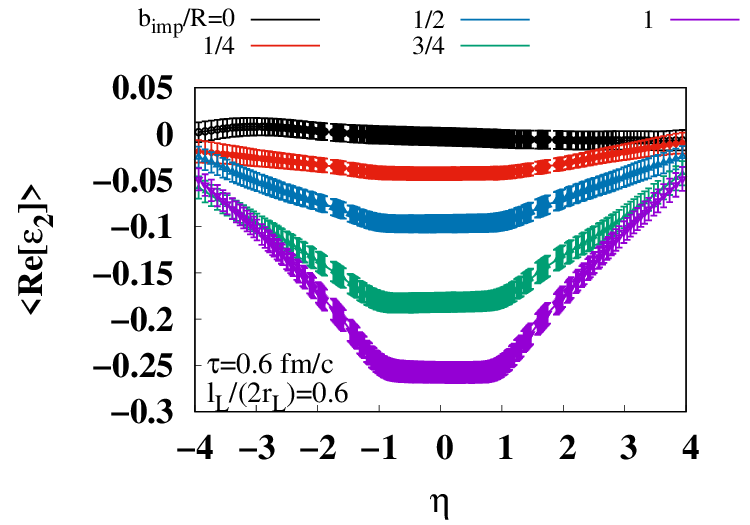}
\end{minipage}%
\hfill
\begin{minipage}{0.45\textwidth}
    \centering
    \includegraphics[width=\textwidth]{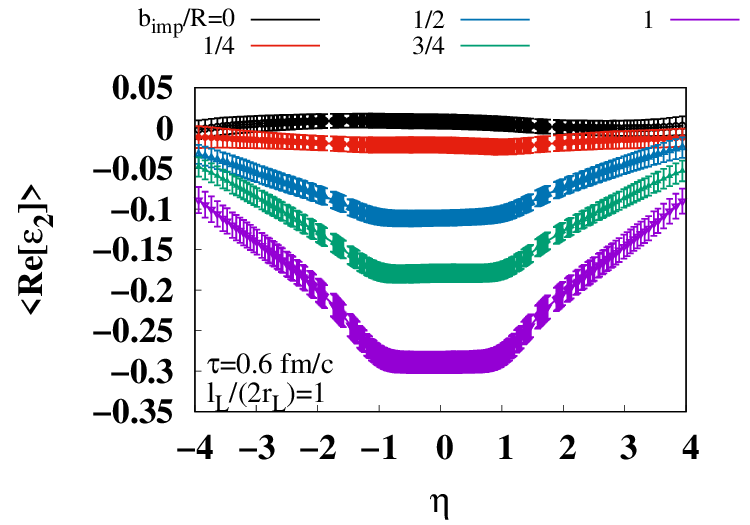}
\end{minipage}
\begin{minipage}{0.45\textwidth}
    \centering
    \includegraphics[width=\textwidth]{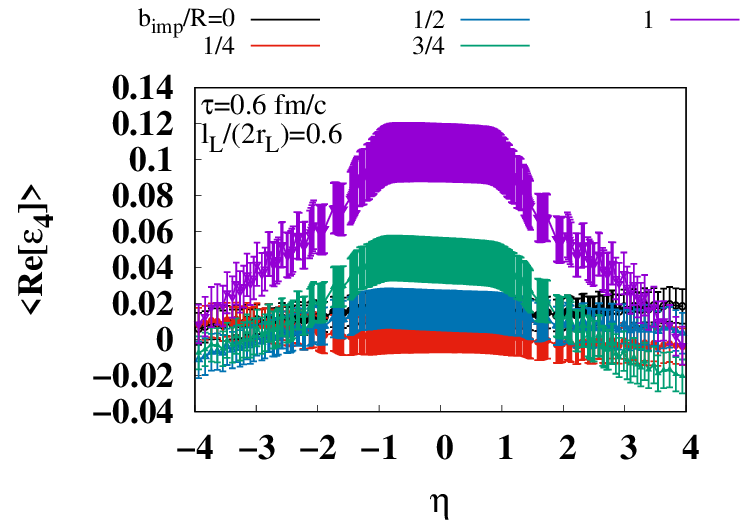}
\end{minipage}%
\hfill
\begin{minipage}{0.45\textwidth}
    \centering
    \includegraphics[width=\textwidth]{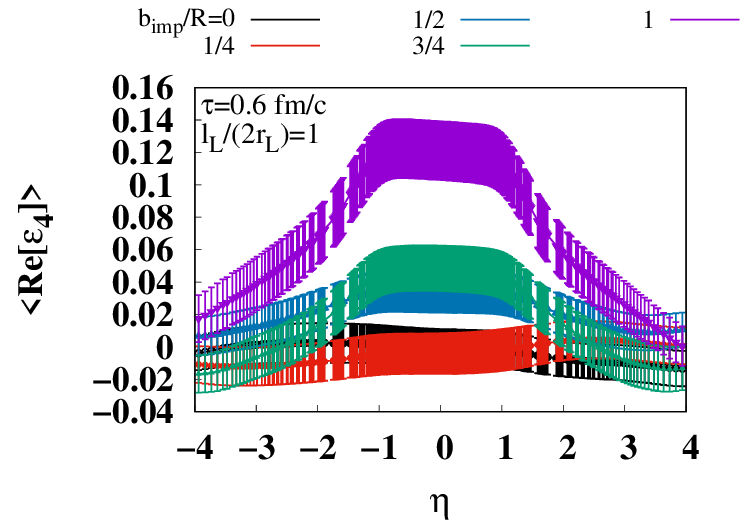}
\end{minipage}
\caption{
The real part of the eccentricity with $n=2$ and $4$, ${\rm Re}\varepsilon_2$ and ${\rm Re}\varepsilon_4$, 
for different values of the longitudinal correlation length, 
$\cl/2\rl=0.6$ and $1$, 
and various impact parameters $\imp/R=0, 1/4, 1/2, 3/4$, and $1$ 
at $\tau=0.6$ fm/c. 
The upper panels show the $n=2$ results, 
with the left and right panels showing $\cl/2\rl=0.6$ and $1$, 
respectively. 
The lower panels show the $n=4$ results, 
with the left and right panels showing $\cl/2\rl=0.6$ and $1$, 
respectively. \
}
 \label{Fig:ecc}
\end{figure*}
In Fig.~\ref{Fig:ecc}, we show the real part of the eccentricity with $n=2$ and $4$, ${\rm Re}\varepsilon_2$ and ${\rm Re}\varepsilon_4$, for different values of the longitudinal correlation length, $\cl/2\rl=0.6$ and $1$, and various impact parameters, $\imp/R=0, 1/4, 1/2, 3/4$ and $1$, at $\tau=0.6$ fm/c.
It is found that both ${\rm Re}\varepsilon_2$ and ${\rm Re}\varepsilon_4$ are consistent with zero within the error in a central collision ($\imp/R = 0$), while, as $\imp$ increases, they become non-zero, indicating the formation of an anisotropic shape of the glasma.
These behaviors of the eccentricities are consistent with the shape of the LRF energy density of the glasma shown in Fig.\ref{Fig:el_tl}.

\begin{figure*}[tp]
\begin{minipage}{0.45\textwidth}
    \centering
    \includegraphics[width=\textwidth]{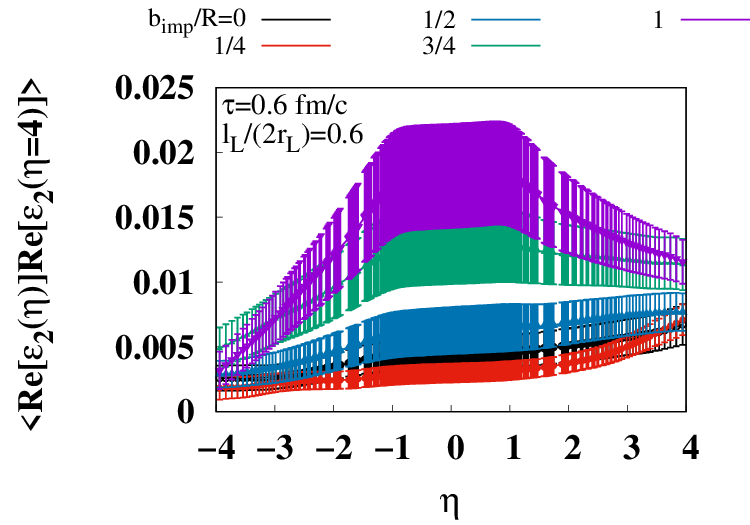}
\end{minipage}%
\hfill
\begin{minipage}{0.45\textwidth}
    \centering
    \includegraphics[width=\textwidth]{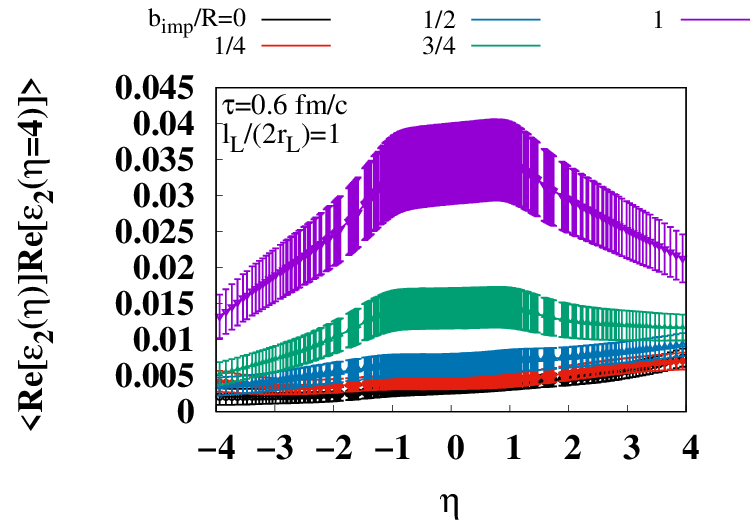}
\end{minipage}
\begin{minipage}{0.45\textwidth}
    \centering
    \includegraphics[width=\textwidth]{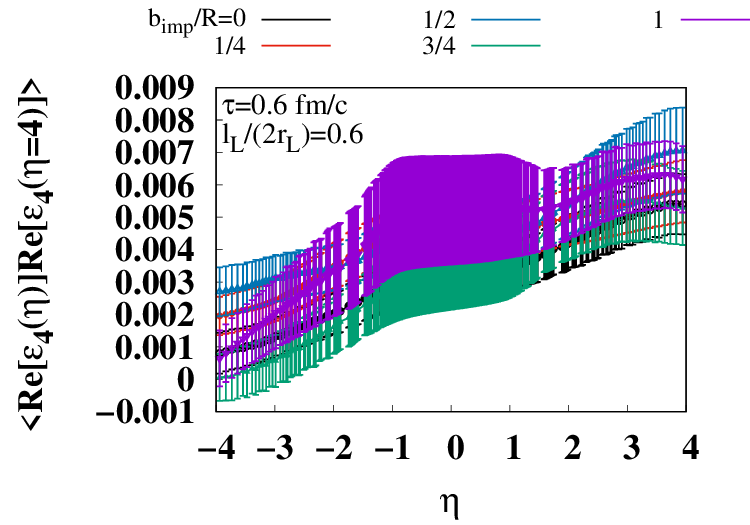}
\end{minipage}%
\hfill
\begin{minipage}{0.45\textwidth}
    \centering
    \includegraphics[width=\textwidth]{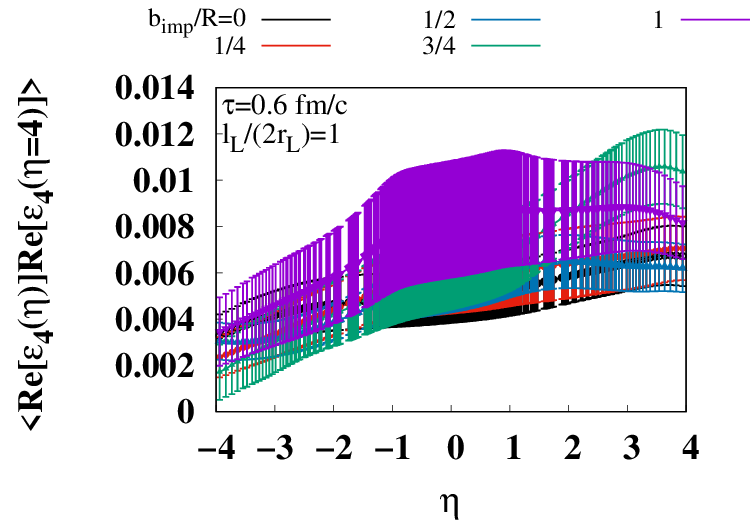}
\end{minipage}
\caption{
The rapditiy correlation of the real part of the eccentricities 
between $\eta$ and $\eta=4$, with $n=2$ and $4$, as defined in 
Eq.~\refeq{Eq:el_col} 
for different values of the longitudinal correlation length, $\cl/2\rl=0.6$ and $1$, 
and various impact parameters, $\imp/R=0, 1/4, 1/2, 3/4$ and $1$, 
at $\tau=0.6$ fm/c. 
The upper panels show the $n=2$ results, 
with the left and right panels showing $\cl/2\rl=0.6$ and $1$, 
respectively. 
The lower panels show the $n=4$ results, 
with the left and right panels showing $\cl/2\rl=0.6$ and $1$, 
respectively.
}
 \label{Fig:el_col}
\end{figure*}
Next, we consider the rapidity correlation of the real part of the eccentricities between $\eta$ and $\eta=4$,
\begin{align}
\average{
{\rm Re} \varepsilon_n(\eta)
{\rm Re} \varepsilon_n(4)}\ .
\label{Eq:el_col}
\end{align}
The eccentricity correlations are expected to be closely related to the decorrelation of flow, which is experimentally observed at LHC~\cite{Dec_flow1,Dec_flow2}.
In Fig.~\ref{Fig:el_col}, we show the rapidity correlation of the real part of the eccentricities between $\eta$ and $\eta=4$, with $n=2$ and $4$, for different values of the longitudinal correlation length, $\cl/2\rl = 0.6$ and $1$, and various impact parameters, $\imp/R = 0, 1/4, 1/2, 3/4$ and $1$, at $\tau = 0.6$ fm/c.
For $\imp/R=0$, the rapidity correlation is found to be a monotonic increasing function of $\eta$, which is expected to reflect the decorrelation effect that increases with the distance between two observed rapidities.
As the impact parameter increases, an enhancement around mid-rapidity gradually emerges, reflecting the increase in the magnitude of the eccentricity as observed in Fig.~\ref{Fig:ecc}.
Thus, the eccentricity correlation are affected by the comparison of this enhancement and the decorrelation effect with increasing rapidity separation.

\subsection{Angular Momentum}\label{Sec:AM}
Here, we discuss the angular momentum of the glasma generated in non-central collisions due to the asymmetry of the geometry with respect to the $x^1$-axis.
This angular momentum, perpendicular to the reaction plane, is expected to be converted into the global spin polarization of observed hadrons~\cite{GlobalPolarization1,GlobalPolarization2}.
The numerical simulations of the generation of angular momentum and vorticity in non-central collisions have been studied using the event-by-event generators such as the Hijing model~\cite{DengHuang_vorticity}, the AMPT model~\cite{Jiang:2016woz,Wei:2018zfb,Xia:2018tes}, and UrQMD model~\cite{Deng:2020ygd}. However, a CGC-based calculation of the angular momentum perpendicular to the reaction plane, focusing on the collective behavior of soft gluons liberated by the collision, is still missing.

\begin{figure*}[tp]
\begin{minipage}{0.45\textwidth}
    \centering
    \includegraphics[width=\textwidth]{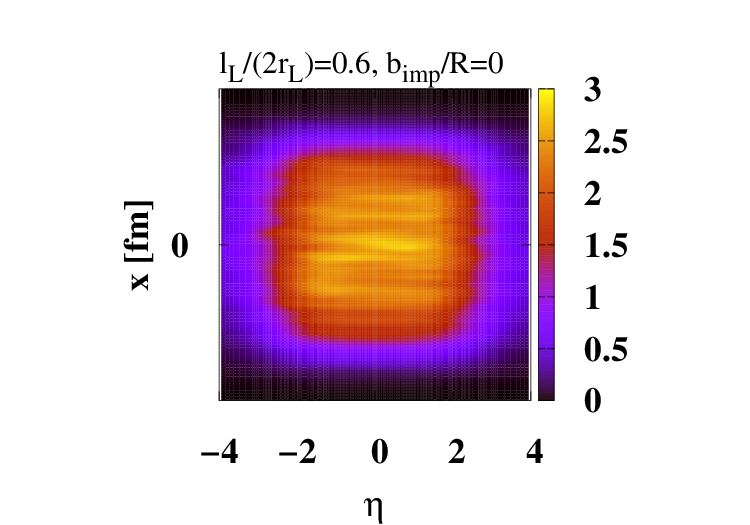}
\end{minipage}%
\hfill
\begin{minipage}{0.45\textwidth}
    \centering
    \includegraphics[width=\textwidth]{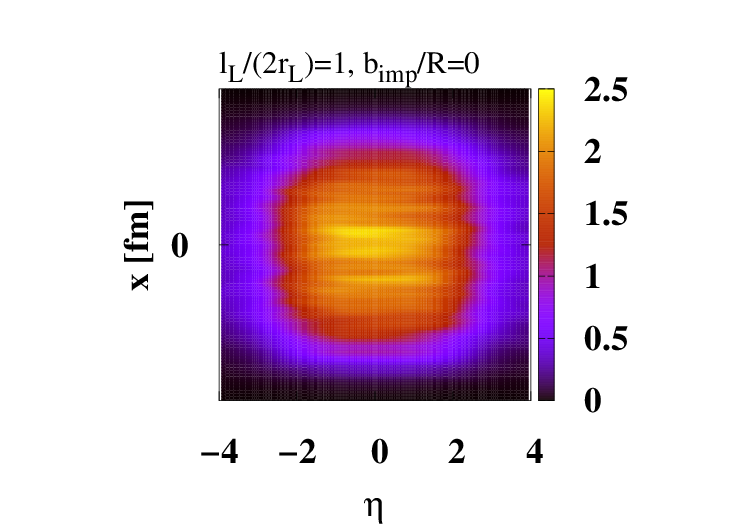}
\end{minipage}
\begin{minipage}{0.45\textwidth}
    \centering
    \includegraphics[width=\textwidth]{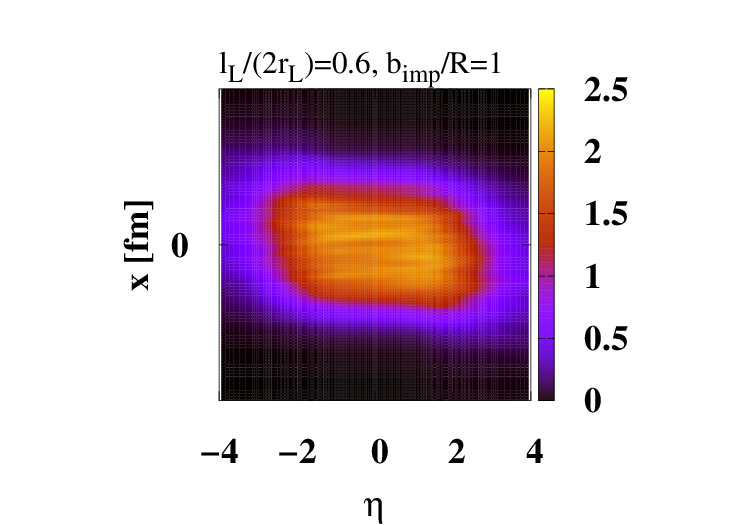}
\end{minipage}%
\hfill
\begin{minipage}{0.45\textwidth}
    \centering
    \includegraphics[width=\textwidth]{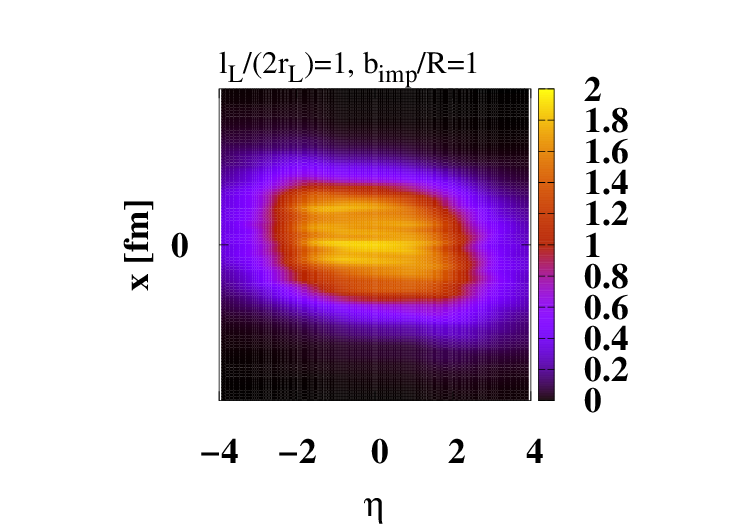}
\end{minipage}
\caption{
The LRF energy density $\elrf$ on the reaction plane ($x^2=0$) 
for two choices of the longitudinal correlation length, 
$\cl/2\rl=0.6$ and $1$, and impact parameter, 
$\imp/R=0$ and $1$ at $\tau=0.6$ fm/c. 
The upper panels show the results for the central collision 
with $\imp/R=0$, 
with the left and right panels showing 
$\cl/2\rl=0.6$ and $1$, respectively. 
The lower panels show the results for the non-central collision 
with $\imp/R=1$, 
with the left and right panels showing $\cl/2\rl=0.6$ and $1$, 
respectively. 
}
 \label{Fig:e_rp}
\end{figure*}
In Fig.~\ref{Fig:e_rp}, we show the LRF energy density $\elrf$ on the reaction plane ($x^2=0$) for two choices of the longitudinal correlation length, $\cl/2\rl=0.6$ and $1$, and impact parameters, $\imp/R=0$ and $1$, at $\tau=0.6$ fm/c.
It is found that while the LRF energy density in central collisions ($\imp/R=0$) is symmetric with respect to the $x^1$-axis, in non-central collisions ($\imp/R=1$), it is distributed diagonally across $x^1=0$ and $\eta=0$.
This result suggests the generation of angular momentum perpendicular to the reaction plane in non-central collisions.

To estimate the angular momentum suggested in Fig.~\ref{Fig:e_rp}, we introduce the rapidity distribution of the angular momentum density perpendicular to the reaction plane as
\begin{align}
\frac{d L^2}{d\eta}
&\equiv \int d^2\bxp \tau \frac{1}{2} \varepsilon^{2ij} M^\tau_{~ij}\nonumber\\
&= \int d^2\bxp \tau \left( x^3 T^{\tau 1} - x^1 T^{\tau 3} \right)\ ,\label{Eq:ang}
\end{align}
where $\tau$ in the integrand stems from the invariant space-time measure in Milne coordinates, and $M^{ijk}=x^j T^{ik} - x^k T^{ij}$ is an angular momentum tensor.
Especially, $M^{\tau ij}$ is a component of the current, which is oriented perpendicular to a hypersurface of constant proper time.
It is noted that, nevertheless, the integration of $\tau M^{\tau ij}$ over the hypersurface, which is an integration of $\frac{d L^2}{d\eta}$ over the rapidity, is not conserved in the $\tau$ evolution.
Later, we focus on the contribution from the Poynting vector parallel to the collision axis and drop $x^3 T^{\tau 1}$ in Eq.~\refeq{Eq:ang}.
The transverse component of the EM tensor that we omit should be tiny compared to the longitudinal component in the early stage.

\begin{figure*}[tp]
\begin{minipage}{0.45\textwidth}
    \centering
    \includegraphics[width=\textwidth]{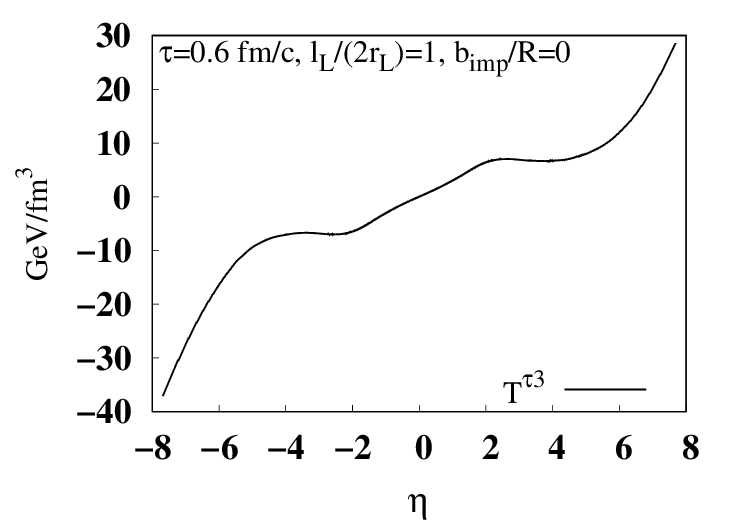}
\end{minipage}%
\hfill
\begin{minipage}{0.45\textwidth}
    \centering
    \includegraphics[width=\textwidth]{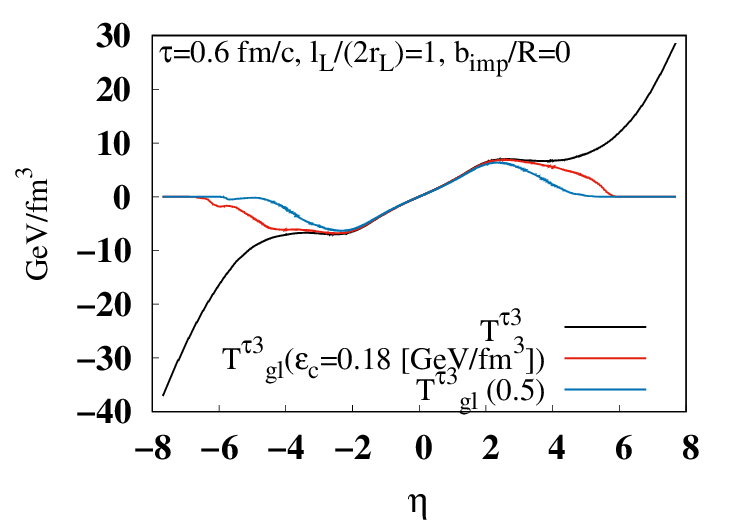}
\end{minipage}
\caption{
Left: the rapidity profile of the longitudinal Poynting vector $T^{\tau 3}$, 
averaged over the transverse plane 
in a significantly wide rapidity region for a single event.
This is presented for $\cl/2\rl=1$ and $\imp/R=0$ at $\tau=0.6$ fm/c.
Right: the rapidity profile of the longitudinal Poynting vector,
the un-subtracted one $T^{\tau 3}$ (the same as that in the left panel) and the newly defined one $T^{\tau 3}_{\rm gl}$ for $\ec=0.18$ and $0.5$ GeV/fm$^3$, 
averaged over the transverse plane 
in a significantly wide rapidity region for a single event.
All the results in left and right panels are obtained from the same single-event simulation.
}
 \label{Fig:ttau3}
\end{figure*}
First, we consider the longitudinal Poynting vector $T^{\tau 3}$, which is directly related to the generation of angular momentum.
In the left panel of Fig.~\ref{Fig:ttau3}, we show the rapidity profile of the longitudinal Poynting vector $T^{\tau 3}$, 
averaged over the transverse plane in a significantly wide rapidity region.
Unlike the other results in this section, this result and the next are obtained from a single event, and the longitudinal size of the lattice and lattice spacing are taken as $L_{\teta}=7168$ and $a_{\teta}=0.0006$, respectively.
It is found that the longitudinal Poynting vector has negative and positive values in the negative and positive rapidity regions, respectively, indicating the expansion of the system along the collision axis around the collision region.
The magnitude is found to increase with $|\eta|$ in the large rapidity region, as well as the energy density $\varepsilon$ and longitudinal pressure $P_\eta$, as shown in Fig.~\ref{Fig:EP}, indicating that the CYM fields exist in the large rapidity region and contribute to some components of the EM tensor.
If this contribution continues to exist in the limit of large rapidity, the integration of the angular momentum cannot be defined as a finite value.

It is important to note that in the current CGC-based method, both the soft gluons in the nucleus and the soft gluons liberated by the collision are collectively represented by CYM fields. Moreover, since mechanisms like string breaking, which clearly separates the nucleus from the produced matter, are not included, the boundary between them is ambiguous. Here, we assume that within the CYM fields, there is a glasma part, which will eventually become QGP, and a non-glasma part, with the majority of the latter consisting of WW fields. Next, focusing on the fact that the LRF energy density of the WW fields is zero, 
while that of the glasma is large, we differentiate between the glasma and non-glasma parts in the CYM fields based 
on the magnitude of the LRF energy density, and define the EM tensor of the glasma part as 
\begin{align}
T^{\mu\nu}_{\rm gl} = \theta(\elrf-\ec) T^{\mu\nu}
\end{align}
where the step function is introduced to exclude the region with small LRF energy density, 
and $\ec$ is a reference energy density introduced to separate the glasma and non-glasma parts, 
which is set to correspond to the crossover region between the confinement and deconfinement phases. 

It should be noted that this method of subtracting the non-glasma part is highly phenomenological. 
Interestingly, using this method, physical quantities that the WW fields do not originally possess, such as longitudinal pressure and topological charge, 
are scarcely affected. 
On the other hand, as we will see below, physical quantities that the WW fields originally possess, such as $T^{\tau 3}$, 
undergo significant subtraction, especially in regions of large rapidity.

In the right panel of Fig.~\ref{Fig:ttau3}, we show the rapidity profile of the longitudinal Poynting vectors, the original one $T^{\tau 3}$ and the newly defined one $T^{\tau 3}_{\rm gl}$, 
averaged over the transverse plane obtained from the same single simulation as the result shown in the left panel of Fig.~\ref{Fig:ttau3}.
The value of the reference energy density is taken as $\ec=0.18$ or $0.5$ GeV/fm$^3$, referring to the energy density in the crossover region, obtained by lattice calculations~\cite{EOS_lattice}.
In the region of $|\eta| < 2$, the newly defined longitudinal Poynting vector $T^{\tau 3}_{\rm gl}$ agrees well with the original one, regardless of the choice of $\ec$.
In the region of $|\eta| > 2$, in contrast to the original one, $T^{\tau 3}_{\rm gl}$ goes to zero as the rapidity becomes larger.
This suppression starts at smaller $\eta$ values as $\ec$ increases.

\begin{figure*}[tp]
\begin{minipage}{0.45\textwidth}
    \centering
    \includegraphics[width=\textwidth]{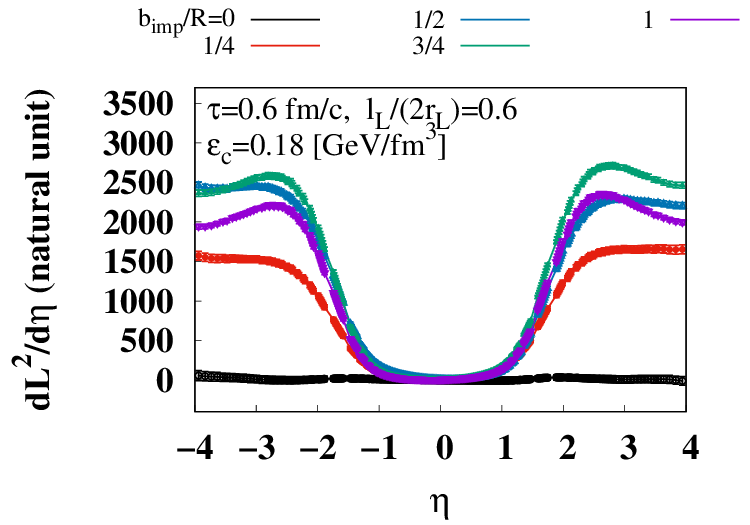}
\end{minipage}%
\hfill
\begin{minipage}{0.45\textwidth}
    \centering
    \includegraphics[width=\textwidth]{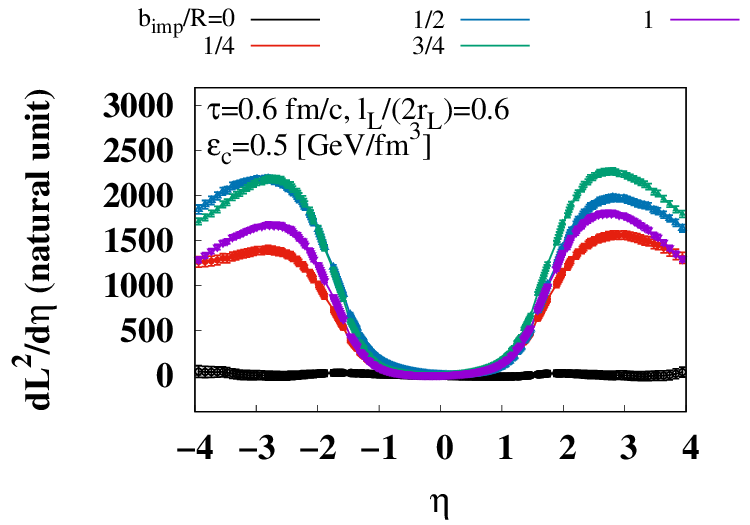}
\end{minipage}
\begin{minipage}{0.45\textwidth}
    \centering
    \includegraphics[width=\textwidth]{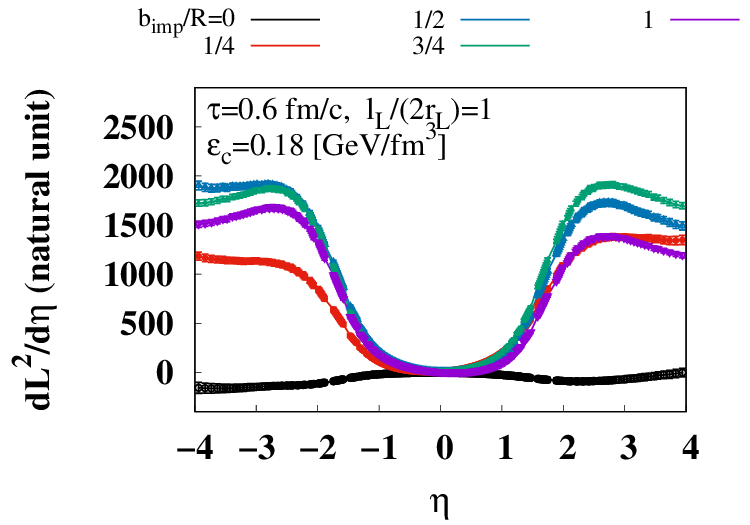}
\end{minipage}%
\hfill
\begin{minipage}{0.45\textwidth}
    \centering
    \includegraphics[width=\textwidth]{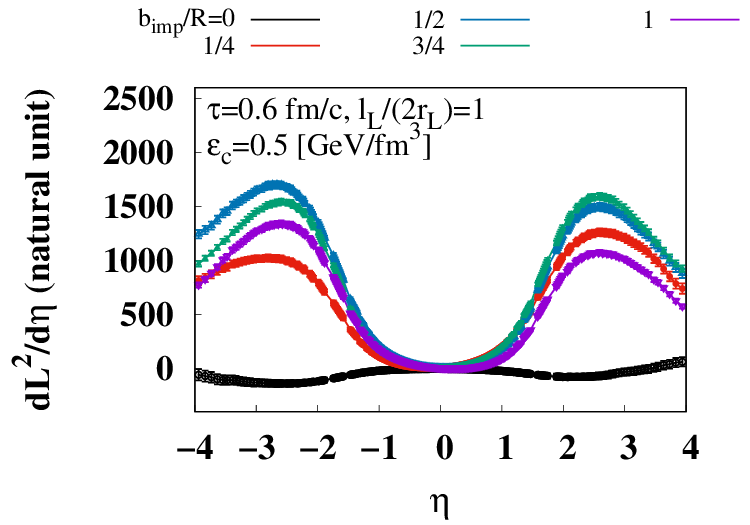}
\end{minipage}
\caption{
The impact parameter dependence of the distribution of the angular momentum perpendicular to the reaction plane, 
defined in Eq.~\refeq{Eq:ang} at $\tau=0.6$ fm/c, 
for two choices of longitudinal correlation length, $\cl/2\rl=0.6$ and $1$, and 
the reference energy density, $\ec=0.18$ and $0.5$ GeV/fm$^3$.
The upper panels show the results for $\cl/2\rl=0.6$ 
with the left and right panels showing 
$\ec=0.18$ and $0.5$ GeV/fm$^3$, respectively. 
The lower panels show the results for $\cl/2\rl=1$  
with the left and right panels showing $\cl/2\rl=0.6$ and $1$, 
respectively. 
}
 \label{Fig:l_1}
\end{figure*}
In Fig.~\ref{Fig:l_1}, 
we show the impact parameter dependence of the distribution of angular momentum perpendicular to the reaction plane, as defined in Eq.~\refeq{Eq:ang}, 
at $\tau=0.6$ fm/c, for two choices of longitudinal correlation length, $\cl/2\rl=0.6$ and $1$, and two reference energy densities, $\ec=0.18$ and $0.5$ GeV/fm$^3$.
The behavior of the results is expected to be reliable enough. However, it is difficult to say that they form the expected symmetric shape with respect to the $x^1 x^2=0$ plane. To obtain more quantitatively accurate results, it may be necessary to perform calculations using a larger number of events.
It is interesting to note that at mid-rapidity, the generation of angular momentum appears to be zero, regardless of the value of the impact parameter.
On the other hand, as the rapidity increases, the generation of angular momentum is observed, and its magnitude strongly depends on the value of the impact parameter.
The peak magnitude appears around $\eta=2-3$, which is close to the peak in the longitudinal component of the Poynting vector shown in the right panel of Fig.~\ref{Fig:ttau3}.
The highest peak appears around $\imp/R=1/2-3/4$, indicating that the angular momentum depends on $\imp$ non-monotonically. Such a behavior is consistent with the phenomenological estimate~\cite{AngularHydro} and numerical calculations based on the HIJING model~\cite{DengHuang_vorticity}.

It is interesting that in the current CGC-based simulation there is little or no angular momentum generation at mid-rapidity, even for non-zero impact parameters. Given that the CGC description is more reliable at higher collision energies, this suggests that little angular momentum would be retained in the mid-rapidity region in the glasma and subsequently in the QGP at very high energies. In fact, the experimentally measured global spin polarization of hadrons at mid-rapidity is indeed tiny at high energies such as $\sqrt{s_{\rm NN}}=200$ GeV~\cite{GlobalPolarization2} and 2.76 TeV and 5.02 TeV~\cite{EXP:polarization1}. In other words, effects beyond the high-energy limit's description may be more important for explaining the spin polarization of hadrons observed at mid-rapidity at lower collision energies~\cite{GlobalPolarization2}. 
This is consistent with experimental results showing that spin polarization is enhanced as the beam energy decreases.

Finally, let us roughly estimate the magnitude of the total angular momentum retained in the glasma. Assuming the angular momentum of the glasma exists in the same rapidity range as that of the longitudinal EM tensor of glasma, i.e., $|\eta|<4$-$6$, as shown in the right panel of Fig.~\ref{Fig:ttau3}, using the results as shown in Fig.~\ref{Fig:l_1}, we obtain that the total angular momentum of the glasma is of the order of $10^4$ for Au-Au with $\imp/R\sim 1/2-1$ at $\sqrt{s_{\rm NN}}=200$ GeV. This is not inconsistent with the result from simulations from, e.g., the HIJING model~\cite{DengHuang_vorticity} and hydrodynamical calculation~\cite{AngularHydro}.

\subsection{Vortex}\label{Sec:AV}
One consequence of the appearance of angular momentum is the formation of vortex in the medium. Although the usual fluid vortex is well defined only when a hydrodynamic picture is adapted (such as in the QGP phase), we could extend this concept to the glasma phase for the purpose of quantifying the local rotation in the glasma. We thus define the vorticity of the glasma as:
\begin{align}
\bm{\omega} \equiv \bm{\nabla} \times \bm{v}\ ,\label{Eq:omega}
\end{align}
where $v^i \equiv p^i/p^0 = T^{0 i}/T^{0 0}$ is the velocity field.
If the system is rotating as a rigid body, the local angular momentum in Minkowski coordinates, $\bm{L}=\bm{x} \times \bm{p}$, is 
related to the vorticity as $\bm{L}=p^0 [\bm{x}^2 \bm{\omega} - (\bm{x} \cdot \bm{\omega}) \bm{x}]/2$.
Thus, the magnitude of the vortex is given by using the component of angular momentum that is parallel to the vortex, $L_\parallel$, as
\begin{align}
\omega = \frac{2 L_\parallel}{p^0 r^2_\omega}\ ,\label{Eq:omega2}
\end{align}
where $r_\omega$ is a distance in the spatial plane perpendicular to the vortex. Since the angular momentum in our simulation is perpendicular to the reaction plane, we thus use $L_\parallel=L^2$ to evaluate the vorticity.

\begin{figure*}[tp]
\begin{minipage}{0.45\textwidth}
    \centering
    \includegraphics[width=\textwidth]{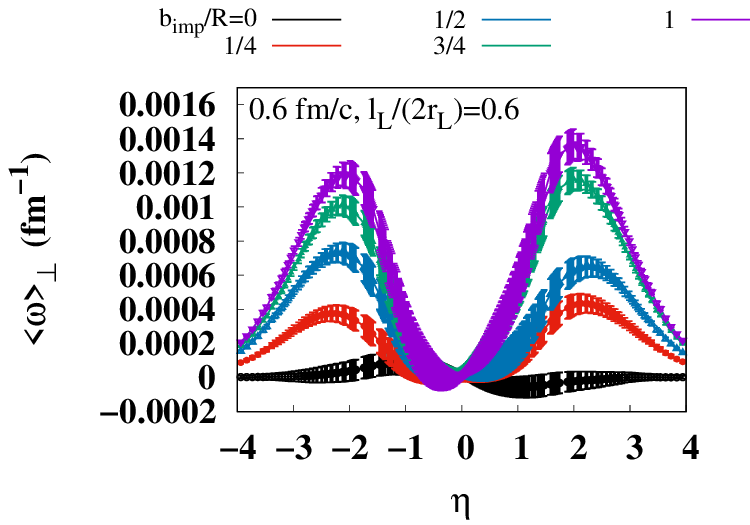}
\end{minipage}%
\hfill
\begin{minipage}{0.45\textwidth}
    \centering
    \includegraphics[width=\textwidth]{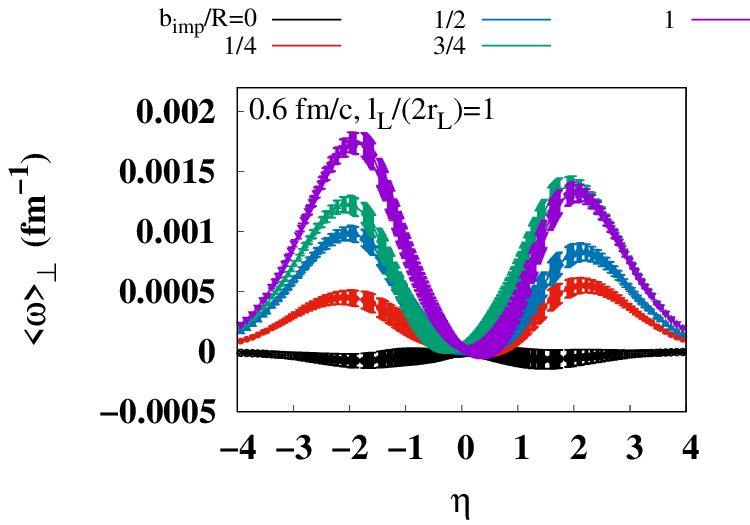}
\end{minipage}
\caption{
The rapidity and impact parameter dependence of the vorticity $\omega$ averaged over transverse plane and over events, as defined in Eq.~\refeq{Eq:spatial}.
This is presented for $\cl/2\rl=0.6$ (left panel) 
and $1$ (right panel) at $\tau=0.6$ fm/c.
}
 \label{Fig:omega}
\end{figure*}
We take the average over transverse plane with a weight by the LRF energy density as done in Ref.~\cite{DengHuang_vorticity}:
\begin{align}
\average{\omega}_\perp =
\frac{\int d^2\bxp \elrf \omega}{\int d^2\bxp \elrf}\ .\label{Eq:spatial}
\end{align}
In Fig.~\ref{Fig:omega}, we show the rapidity and impact parameter dependence of the vorticity $\omega$, averaged over both the transverse plane and events. This is presented for different values of the longitudinal correlation length, $\cl/2\rl = 0.6$ and $1$, at $\tau=0.6$ fm/c.
The vorticity also shows two peaks, similar to the angular momentum distribution.
However, the interval between the peaks is narrower compared to the angular momentum distribution, and the decay rate in the regions with larger rapidity than the peaks is much faster.
This behavior is expected to be due to the fact that the vorticity contains a term proportional to $r^{-2}_\omega$ as shown in Eq.~\refeq{Eq:omega2}.
Additionally, unlike the angular momentum distribution, at least in the region of $\imp \leq R$, the magnitude of the vorticity becomes larger as the impact parameter increases.
This suggests that in this impact parameter region, the local rotation of the generated matter becomes faster as the impact parameter increases.

\section{Summary}\label{Sec:6}
In this study, we have applied the 3D glasma simulation method using Milne coordinates to investigate the early stages of the central and non-central collisions of Au-Au at $\snn = 200$ GeV. 
The 3D glasma simulation accounts for the longitudinal structure of the nucleus, the finite thickness of nucleons and their random positions along the collision axis, which were missing in the conventional boost-invariant glasma description.
Consequently, it becomes possible to extend the description of the glasma, limited to mid-rapidity under the boost invariance, to a larger rapidity region.

We have constructed the initial conditions for the pre-collision nuclei as a simple extension of that for the boost-invariant IP-glasma model.
We have adjusted the MV model parameter $g^2\mu_{\rm 2D}$ in the pre-collision nucleus model to reproduce the saturation scale of the nucleus estimated by the IP-sat model~\cite{IPsat_A}.

For central collisions, we have analyzed the rapidity profiles of the LRF energy density first. 
Through the comparison between the LRF energy density of the glasma and experimental data of the particle-rapidity profile of hadrons, we find the better choices of the longitudinal correlation length of the classical color charge density, a model parameter.
Moreover, the LRF energy density of the glasma at mid-rapidity is found to be close to that estimated by boost-invariant hydrodynamic simulations~\cite{Hydro_for_HIC} and that estimated by the boost-invariant glasma simulation with MV model~\cite{KNV_energy}.

Then, we have calculated the rapidity profiles of the energy density, transverse and longitudinal pressures, as well as the LRF energy density.
As observed in Ref.~\cite{3Dglasma_FU1,3Dglasma_CPIC,3Dglasma_SS,3Dglasma_FU2,3Dglasma_analytic2}, 
we have found that the longitudinal pressure and energy density extend to larger rapidity regions and even increase in larger rapidity regions, 
while the transverse pressure and LRF energy density are localized near central rapidity.
The discrepancy between the energy density and the LRF energy density indicates that Bjorken flow no longer accurately describes the longitudinal flow of glasma in the large rapidity region.

We have studied the rapidity profile of the squared topological charge and the transverse correlation of the axial charge in the glasma.
We have found that the transverse correlation functions at mid-rapidity decrease in the region $\qs r_\perp < 3$, and then, at $\qs r_\perp > 3$, their values are consistent with $0$ within the error.
Such the behavior of the transverse correlation function is found not to change under proper time evolution.
These results are in qualitative agreement with previous studies based on the boost-invariant glasma with the MV model~\cite{Ruggieri_topological}.

For non-central collisions, we have studied quantities that depend on the impact parameter. 
First, we have examined the eccentricity, a measure of the spatial anisotropy of the system in the transverse plane perpendicular to the collision axis. 
It is found that the real part of eccentricity with $n=2$ and $4$ are consistent with zero within the error in a central collision, while, as the impact parameter increases, they become non-zero values, especially for the near mid-rapidity region, indicating the formation of an anisotropic shape of the glasma.
The rapidity correlation of the real part of eccentricity with $n=2$ and $4$ is formed by the comparison of the decorrelation effect from the distance and the enhancement of the eccentricity around mid-rapidity~\cite{AngularHydro}.

Then, we have studied the generation of the angular momentum of the glasma, which is caused by the asymmetry of the glasma's geometry with respect to the impact parameter direction.
To estimate the angular momentum carried solely by the glasma, 
we have subtracted the contribution from the small LRF energy density part of the CYM field, 
which cannot be regarded as the glasma that will eventually evolve into the QGP.
We have confirmed that the angular momentum is generated due to the asymmetric geometry of the glasma.
Although it had been based on a very rough estimate, the angular momentum of the glasma had been found to be in the same numerical range as the phenomenological estimate made using hydrodynamics or parton cascade models. 

Finally, we have also confirmed the generation of the vortex of the glasma with a non-zero impact parameter, which indicates the glasma generated in non-central collisions to be locally rotating. The vortex of the generated matter has been found to become larger as the impact parameter increases.

The effects not considered in the current calculations, as well as making the model parameters more realistic, 
should be contemplated to enable more quantitative discussions. For example, determining the saturation scale based on the IP-sat model in a self-consistent manner or changing the number of colors from 2 to 3 would fall into these categories.

Additionally, the next step, which is actually being explored currently, is to use the 3D glasma description as an initial state model to provide the initial conditions for hydrodynamics and make comparisons with experimental data. This will allow us to evaluate the extent to which incorporating the effects beyond the shock-wave approximation contributes to reproducing the rapidity-dependent behavior of physical observables.

Furthermore, when providing the initial conditions for hydrodynamics, it is necessary to extract local flow and fluid properties from the EM tensor of the CYM field estimated on the lattice. In this case, more serious consideration should be given to the errors arising from the discretization of the EM tensor on the lattice, compared to when considering bulk quantities as focused on in the current paper. The recently proposed improved definition of the Yang-Mills field's EM tensor on the lattice may help address such issues~\cite{EMtensor_lat}.

\section*{Acknowledgments}
We are grateful to Y. Hidaka and K. Watanabe for helpful communications and discussions.
This work is supported by the Natural Science Foundation of Shanghai (Grant No. 23JC1400200), the National Natural Science Foundation of China (Grant No. 12225502, No. 12075061, and  No. 12147101), and the National Key Research and Development Program of China (Grant No. 2022YFA1604900). 
We acknowledge Yukawa-21 for providing the computational resources.




\begin{thebibliography}{99}

\bibitem{3Dglasma_FU1}
H. Matsuda, and X.-G. Huang, Phys. Rev. D \textbf{108}, 114008 (2023).

\bibitem{Hydro_for_HIC}
U. W. Heinz, and P. F. Kolb,
Nucl. Phys. A {\textbf 702}, 269 (2002).

\bibitem{UrQMD1998}
M. Bleicher {\it et al}. J. Phys. G \textbf{25}, 1859-1896 (1999).
\bibitem{UrQMD1999}
S. A. Bass {\it et al}. Prog. Part. Nucl. Phys. \textbf{41}, 255-369 (1998).

\bibitem{JAM}
Y. Nara, N. Otuka, A. Ohnishi, K. Niita, and S. Chiba, Phys. Rev.C {\textbf 61}, 024901 (2000).


%
%
%
%


\bibitem{MonteCarlo}
M. L. Miller, K. Reygers, S. J. Sanders, and P. Steinberg, 
Ann. Rev. Nucl. Part. Sci. \textbf{57}, 205-243 (2007).

\bibitem{TRENTo1}
J. S. Moreland, J. E. Bernhard, and S. A. Bass, 
Phys. Rev. C. \textbf{92}, 011901 (2015).
\bibitem{TRENTo2}
W. Ke, J. S. Moreland, J. E. Bernhard, and S. A. Bass,
Phys. Rev. C. \textbf{96}, 044912 (2017).

\bibitem{IPglasma}
B. Schenke, P. Tribedy, and R. Venugopalan,
Phys. Rev. Lett. {\textbf 108}, 252301 (2012)
Phys. Rev. C {\textbf  86}, 034908 (2012).

\bibitem{MV}
L. D. McLerran, and R. Venugopalan, 
Phys. Rev. D {\textbf 49}, 2233--2241 (1994);
Phys. Rev. D {\textbf 49}, 3352--3355 (1994);
Phys. Rev. D {\textbf 50}, 2225--2233 (1994).
\bibitem{CGCreview}
E. Iancu, and R. Venugopalan, arXiv:0303204.


\bibitem{MVcollision}
A. Kovner, L. D. McLerran and H. Weigert, Phys. Rev. D \textbf{52}, 6231-6237 (1995).


\bibitem{2Dglasma_KV}
A. Krasnitz, and R. Venugopalan, Nucl. Phys. B \textbf{557}, 237 (1999); Phys. Rev. Lett. \textbf{84}, 4309-4312 (2000).
\bibitem{KNV_SU3}
A. Krasnitz, Y. Nara, and R. Venugopalan, 
Phys. ReV. Lett. {\textbf 87}, 192302 (2001):
Nucl. Phys. A {\textbf 717}, 268 (2003).
\bibitem{KNV_energy}
A. Krasnitz, Y. Nara, and R. Venugopalan, 
Nucl. Phys. A {\textbf 727}, 427 (2003).
\bibitem{2Dglasma_L}
T. Lappi, Phys. Rev. C \textbf{67}, 054903 (2003).
\bibitem{LM}
T. Lappi and L. McLerran, Nucl. Phys. A\textbf{772}, 200-212  (2006).
\bibitem{InstabilityCYM2006_RV}
P. Romatschke, and R. Venugopalan, Phys. Rev. Lett. \textbf{96}, 062302 (2006); 
Eur. Phys. J. A \textbf{29}, 71 (2006); 
Phys. Rev. D \textbf{74}, 045011 (2006).
\bibitem{InstabilityCYM2008}
J. Berges, S. Scheffler, and D. Sexty, Phys. Rev. D \textbf{77}, 034504 (2008).
\bibitem{InstabilityCYM2009}
J. Berges, D. Gelfand, S. Scheffler, and D. Sexty, Phys. Lett. B \textbf{677}, 210 (2009).
\bibitem{InstabilityCYM2012FG}
K. Fukushima, and F. Gelis, Nucl. Phys. A \textbf{874}, 108 (2012).
\bibitem{InstabilityCYM2012BSSS}
J. Berges, S. Scheffler, S. Schlichting, and D. Sexty, Phys. Rev. D \textbf{85}, 034507 (2012).
\bibitem{InstabilityCYM2013}
J. Berges, and S. Schlichting, Phys. Rev. D \textbf{87}, 014026 (2013).
\bibitem{PressureEG2013}
T. Epelbaum, and F. Gelis, Phys. Rev. Lett. \textbf{111}, 232301 (2013).
\bibitem{Turbulent2014}
J. Berges, K. Boguslavski, S. Schlichting, and R. Venugopalan, Phys. Rev. D \textbf{89}, 074011 (2014).
\bibitem{Universal2014}
J. Berges, K. Boguslavski, S. Schlichting, and R. Venugopalan, Phys. Rev. D \textbf{89}, 114007 (2014).
\bibitem{Entropy_kyoto}
H. Matsuda, T. Kunihiro, A. Ohnishi, and T. T. Takahashi, PTEP \textbf{2022}, 073D02 (2022).
\bibitem{SmallTime_CCM}
M. E. Carrington, A. Czajka, and S. Mr$\acute{\rm o}$wczy$\acute{\rm n}$ski, 
Phys. Rev. C {\textbf 106}, 3, 034904 (2022).
\bibitem{SmallTime_CCFMP}
M. E. Carrington, W. N. Cowie, B. T. Friesen, S. Mr$\acute{\rm o}$wczy$\acute{\rm n}$ski, and D. Pickering, 
Phys. Rev. C {\textbf 108}, 5, 054903 (2023).
\bibitem{SmallTime_CM}
M. E. Carrington, S. Mr$\acute{\rm o}$wczy$\acute{\rm n}$ski, and J.-Y. Ollitrault, arXiv:2406.14463.
\bibitem{HP_CPIC}
D. Avramescu, V. Greco, T. Lappi, H. M$\ddot{\rm a}$ntysaari, and David M$\ddot{\rm u}$ller, arXiv:2409.10564; arXiv:2409.10565.


\bibitem{Dec_flow1}
V. Khachatryan {\it et al}. [CMS Collaboration], Phys. Rev. C {\textbf 92}, 034911 (2015). 
\bibitem{Dec_flow2}
M. Aaboud {\it et al.} [ATLAS Collaboration], Eur. Phys. J. C {\textbf 78}, 142 (2018).
\bibitem{Dec_tra1}
H. Agakishiev {\it et al.} [STAR Collaboration], Phys. Lett. B {\textbf 704}, 467-473 (2011).
\bibitem{Dec_tra2}
S. Acharya {\it et al.} [ATLAS Collaboration], Phys. Lett. B {\textbf 804}, 135375 (2020).
\bibitem{Correlation_from_hydro_init_to_obs}
O. Savchuk, arXiv:2402.12504.




\bibitem{JIMWLK1}
J. J.-Marian, A. Kovner, A. Leonidov, and H. Weigert,
Nucl. Phys. B {\textbf 504}, 415--431, (1997);
Phys. Rev.  D {\textbf 59},  014014 (1998);
Phys. Rev.  D {\textbf 59},  034007 (1999);
Phys. Rev.  D {\textbf 59},  09990  (1999) [Erratum].
\bibitem{JIMWLK2}
Jalilian-Marian, J.; Kovner, A.; Weigert, H. 
{\em Phys. Rev. D} {\textbf 1999}, {\em 59}, 014015.
\bibitem{JIMWLK3}
E. Iancu, A. Leonidov, and L. D. McLerran, 
Nucl. Phys. A {\textbf 692}, 583--645 (2001).
\bibitem{JIMWLK4}
E. Ferreiro, E. Iancu, A. Leonidov, and L. D. McLerran,
Nucl. Phys. A {\textbf 703}, 489--538 (2002).
\bibitem{JIMWLK5}
E. Iancu, A. Leonidov, and L. D. McLerran, 
Phys. Lett. B {\textbf 510}, 133--144 (2001).
\bibitem{JIMWLK6}
E. Iancu, and L. D. McLerran,
Phys. Lett. B {\textbf 510}, 145--154 (2001).


\bibitem{3DIPglasma_JIMWLK1}
B. Schenke and S. Schlichting, Phys. Rev. C \textbf{94}, 044907 (2016).
\bibitem{3DIPglasma_JIMWLK2}
B. Schenke, S. Schlichting, and P. Singh, 
Phys. Rev. D. {\textbf 105}, 094023 (2022).
\bibitem{3DIPglasma_JIMWLK3}
S. McDonald, S. Jeon, S, and G. Gale. 
Phys. ReV. C. {\textbf 108}, 064910 (2023).


\bibitem{3Dglasma_CPIC}
D. Gelfand, A. Ipp and D. I. M$\ddot{\rm u}$ller, Phys. Rev. D \textbf{94}, 014020 (2016).
\bibitem{3Dglasma_SS}
S. Schlichting and P. Singh, Phys. Rev. D \textbf{103}, 1, 014003 (2021).


\bibitem{3Dglasma_analytic1}
A. Ipp, D. I. M$\ddot{\rm u}$ller, S. Schlichting and P. Singh, 
Phys. Rev. D \textbf{104}, 11, 114040 (2021).
\bibitem{3Dglasma_analytic2}
A. Ipp, M. Leuthner, D. I. M$\ddot{\rm u}$ller, S. Schlichting, K. Schmidt, and P. Singh, 
Phys. Rev. D {\textbf 109}, 094040 (2024).


\bibitem{SU2vsSU3}
O. Philipsen, B. Wagenbach, and S. Zafeiropoulos, 
Eur. Phys. J. C {\textbf 79}, 286 (2019).


\bibitem{KV}
Y. V. Kovchegov, Phys. Rev. D {\textbf 54}, 5463--5469 (1996).


\bibitem{JIWMLK_fact}
F. Gelis, T. Lappi, and R. Venugopalan, 
Phys. Rev. D. {\textbf 78}, 054019 (2008);
Phys. Rev. D. {\textbf 78}, 054020 (2008);
Phys. Rev. D. {\textbf 79}, 094017 (2009).


\bibitem{WW_QED}
E. J. Williams, Kong. Dan. Vid. Sel. Mat. Fys. Med. 13N4, 4, 1 (1935).


\bibitem{IPsat1}
J. Bartels, K. Golec-Biernat, and H. Kowalski,
Phys. Rev. D {\textbf 66}, 014001 (2002).
\bibitem{IPsat2}
H. Kowalski, and D. Teaney,
Phys. Rev. D {\textbf 68}, 114005 (2003).

\bibitem{3Dglasma_FU2}
H. Matsuda, and X.-G. Huang, Entropy {\textbf 26}, 167 (2024).


\bibitem{Lappi}
T. Lappi, Eur. Phys. J. C {\textbf 55}, 285 (2008).


\bibitem{IPsat_HERA}
A. H. Rezaeian, M. Siddikov, M. Van de Klundert, and R. Venugopalan, 
Phys. Rev. D {\textbf 87} 034002 (2013).


\bibitem{IPsat_A}
H. Kowalski, T. Lappi, and R. Venugopalan
Phys. Rev. Lett. {\textbf 100}, 022303 (2008).


\bibitem{FG}
K. Fukushima, and F. Gelis, Nucl. Phys. A \textbf{874}, 108 (2012).


\bibitem{RHIC_pion}
I. Bearden {\it et al}. [BRAHMS Collaboration], Phys. Rev. Lett. {\textbf 94}, 162301 (2005).


\bibitem{RHIC_charged_hadron}
B. B. Back {\it et al}. [PHOBOS collaboration], Phys. Rev. Lett. {\textbf 91}, 052303 (2003).





\bibitem{ABJ1}
S. L. Adler, Phys. Rev. {\textbf 177}, 2426 (1969).
\bibitem{ABJ2}
J. S. Bell, and R. Jackiw, Nuovo Cimento A {\textbf 60}, 47 (1969).


\bibitem{CME1}
D. E. Kharzeev, L. D. McLerran, and H. J. Warringa, Nucl. Phys. A {\textbf 803}, 227 (2008).
\bibitem{CME2}
K. Fukushima, D. E. Kharzeev, and H J. Warringa, Phys. Rev. D {\textbf 78}, 074033 (2008).


\bibitem{KKVtopological}
D. Kharzeev, A. Krasnitz, and R. Venugopalan, Phys. Lett. B {\textbf 545}  298-306 (2002).
\bibitem{SphaleronTransition}
M. Mace, S, Schlichting, and R. Venugopalan, Phys Rev D {\textbf 93}, 074036 (2016).
\bibitem{Ruggieri_topological}
M. R. Jia, J. H. Liu, H. F. Zhang, and M. Ruggieri, Phys. Rev. D {\textbf 103} 014026 (2001).


\bibitem{GlobalPolarization1}
Z.-T. Liang, and X .-N. Wang, Phys. Rev. Lett. {\textbf 94}, 102301 (2005).
\bibitem{GlobalPolarization2}
L.~Adamczyk \textit{et al.} [STAR],
Nature \textbf{548}, 62-65 (2017).

\bibitem{DengHuang_vorticity}
W.-T. Deng, and X.-G. Huang, Phys. Rev. C {\textbf 93}, 064907 (2016).
\bibitem{Jiang:2016woz}
Y.~Jiang, Z.~W.~Lin and J.~Liao,
Phys. Rev. C \textbf{94}, no.4, 044910 (2016)
[erratum: Phys. Rev. C \textbf{95}, no.4, 049904 (2017)].
\bibitem{Wei:2018zfb}
D.~X.~Wei, W.~T.~Deng and X.~G.~Huang,
Phys. Rev. C \textbf{99}, no.1, 014905 (2019).
\bibitem{Xia:2018tes}
X.~L.~Xia, H.~Li, Z.~B.~Tang and Q.~Wang,
Phys. Rev. C \textbf{98}, 024905 (2018).
\bibitem{Deng:2020ygd}
X.~G.~Deng, X.~G.~Huang, Y.~G.~Ma and S.~Zhang,
Phys. Rev. C \textbf{101}, 064908 (2020).

\bibitem{EOS_lattice}
A. Bazavov, {\it et al.} [HotQCD Collaboration], Phys. Rev. D {\textbf 90} 094503 (2014).


\bibitem{AngularHydro}
F. Becattini, G. Inghirami, V. Rolando, A. Beraudo, L. Del Zanna, A. De Pace, M. Nardi, G. Pagliara, and V. Chandra, 
Eur. Phys. J. C {\textbf 75}, 406 (2015);
Eur. Phys. J. C {\textbf 78}, 354 (2018) [Erratum].

\bibitem{EXP:polarization1}
S.~Acharya \textit{et al.} [ALICE],
Phys. Rev. C \textbf{101}, 044611 (2020)
[erratum: Phys. Rev. C \textbf{105}, 029902 (2022)].

\bibitem{EMtensor_lat}
K. Boguslavski, T. Lappi, J. Peuron, and P. Singh, Eur. Phys. J. C {\textbf 84}, 4, 368 (2024).


\end{thebibliography}
\end{document}